\documentclass[11pt]{article}
\usepackage{amsfonts}
 
\def\pa{\partial}

\def\g{\gamma} \def\G{\Gamma} %\mbox{\boldmath $A$}
\def\a{\alpha}
\def\b{\beta}
\def\d{\delta} 
\def\e{\epsilon}

\def\l{\lambda} 
\def\m{\mu}
\def\n{\nu}

\def\s{\sigma} \def\S{\Sigma}
\def\t{\tau}

\def\mn{{\mu\nu}}
\def\ab{{\alpha\beta}}
\def\be{\begin{equation}}
\def\ee{\end{equation}}

\def\fg{{\mathfrak g}}
\def\fT{{\mathfrak T}}
\def\fF{{\mathfrak F}}
\def\bea{\begin{eqnarray}}
\def\eea{\end{eqnarray}}
\begin{document}

{\small {\bf Abstract:} This article---summarizing the authors' then
novel formulation of General Relativity---appeared as Chapter 7,
pp.227--264, in {\it Gravitation: an introduction to current
research}\/, L. Witten, ed. (Wiley, New York, 1962), now long out of
print. Intentionally unretouched, this posting is intended to provide
contemporary accessibility to the flavor of the original ideas. Some
typographical corrections have been made: footnote and page numbering
have changed---but not section nor equation numbering, etc. 
Current institutional affiliations are encoded in: 
arnowitt@physics.tamu.edu~, deser@brandeis.edu~,
misner@physics.umd.edu~.}
\vspace{0.5 in}

\renewcommand{\thefootnote}{\roman{footnote}}

\begin{center}\textbf{
\it\LARGE
R.\ Arnowitt,\/\footnote{Physics Department, Syracuse University,
Syracuse, NY. (Research supported in part by the National Science
Foundation and the United States Air Force---Aeronautical Research
Laboratory, WADD, and Office of Scientific Research.)}
S.\ Deser,\/\footnote{Physics Department, Brandeis University, Waltham, MA.
(Research supported in part by the National Science Foundation and the
United States Air Force Office of Scientific Research.)}
and
C.\ W.\ Misner\/\footnote{Palmer Physical Laboratory, Princeton University,
Princeton, NJ. (Alfred P. Sloan Research Fellow.)} 
}
\end{center}
\vspace{0.5 in}

{\Huge\bf\noindent The Dynamics\\
          of General Relativity}
\vspace{1 in}

\noindent{\large\bf 1. Introduction}%
\footnote{The work discussed in this chapter is based on recent research
by the authors. In the text, the original papers will be denoted by
Roman numerals, as given in the references.}

The general coordinate invariance underlying the theory of
relativity creates basic problems in the analysis of the dynamics
of the gravitational field.  Usually, specification of the field
amplitudes and their first time derivatives initially is
appropriate to determine the time development of a field viewed as
a dynamical entity. For general relativity, however, the metric
field $g_\mn$ may be modified at any later time simply by carrying
out a general coordinate transformation.  Such an operation does
not involve any observable changes in the physics, since it merely
corresponds to a relabeling under which the theory is invariant.
Thus it is necessary that the metric field be separated into the
parts carrying the true dynamical information and those parts
characterizing the coordinate system. In this respect, the general
theory is analogous to electromagnetic theory.  In particular, the
coordinate invariance plays a role similar to the gauge invariance
of the Maxwell field.  In the latter case, this gauge invariance
also produces difficulties in separating out the independent
dynamical modes, although the linearity here does simplify the
analysis.  In both cases, the effect of invariance properties
(both Lorentz and ``gauge" invariance) is to introduce redundant
variables in the original formulation of the theory to insure that
the correct transformation properties are maintained.  It is this
clash with the smaller number of variables needed to describe the
dynamics ({\it i.e.}, the number of {\it independent} Cauchy data)
that creates the difficulties in the analysis. In Lorentz
covariant field theories, general techniques (Schwinger, 1951,
1953) (valid both in the quantum and classical domains) have been
developed to enable one to disentangle the dynamical from the
gauge variables. We will see here that, while general relativity
possesses certain unique aspects not found in other theories,
these same methods may be applied. Two important advantages are
obtained by proceeding in this fashion.  First, the physics of
Lorentz covariant field theory is well understood.  Consequently,
techniques which relate back to this area of knowledge will help
one to comprehend better the physics of general relativity.
Second, insofar as quantization of the theory is concerned, a
formulation closely associated with general quantization
techniques which include consistency criteria ({\it i.e.}, the
Schwinger action principle) will be more appropriate for this
highly non-linear theory.  Direct correspondence principle
quantization (which is suitable for linear theories without
constraints) may well prove inadequate here.

A precise determination of the independent dynamical modes of the
gravitational field is arrived at when the theory has been cast
into canonical form and consequently involves the minimal number
of variables specifying the state of the system. At this level,
one will have all the relevant information about the field's
behavior in familiar form. The canonical formalism, involving only
the minimal set of variables (which will turn out to be four), is
also essential to the quantization program, since it yields
directly simple Poisson bracket (P.B.) relations among these
conjugate, unconstrained, variables. Two essential aspects of
canonical form are: (1) that the field equations are of first
order in the time derivatives; and (2) that time has been singled
out so that the theory has been recast into 3+1 dimensional form.
These two features are characteristic of Hamilton (or P.B.)
equations of motion, in contrast to the Lagrange equations.  The
first requirement may be achieved in general relativity, since its
Lagrangian may be written in a form linear in the time derivatives
(which is called the Palatini form).  The type of variable
fulfilling the second requirement is dictated by the desire for
canonical form, and will be seen also to possess a natural
geometrical interpretation.

The use of the Palatini Lagrangian and of 3+1 dimensional notation
does not, of course, impair the general covariance of the theory
under arbitrary coordinate transformations.  In possessing this
covariance, general relativity is precisely analogous to the
parameterized form of mechanics in which the Hamiltonian and the
time are introduced as a conjugate pair of variables of a new
degree of freedom.  When in parameterized form, a theory is
invariant under an arbitrary re-parameterization, just as general
relativity is invariant under an arbitrary change of coordinates.
The action of general relativity will thus be seen to be in
``already parameterized" form.  The well-known relations between
the usual canonical form and the parameter description will thus
provide the key for deriving the desired canonical form for the
gravitational field.  We will therefore begin, in Section 2, with
a brief review of parameterized particle mechanics.  In Section 3,
the Lagrangian of general relativity will be cast into Palatini
and 3+1 dimensional form, and the geometrical significance of the
variables will be discussed.  We will see then that relativity has
a form identical to parameterized mechanics. Section 4 completes
the analysis, to obtain the canonical variables and their
relations as well as the P.B. equations of motion.

Once canonical form is reached, the physical interpretation of
quantities involved follows directly as in other branches of
physics. Thus, the canonical variables themselves represent the
independent excitations of the field (and hence provide the basis
for defining gravitational radiation in a coordinate-independent
way). Further, the numerical value of the Hamiltonian for a
particular state of the system provides the primary definition of
total energy (a definition which amounts to comparing the
asymptotic form of the spatial part of the metric with that of the
exterior Schwarzschild solution).  Similarly, the total momentum
is defined from the generator of {\it spatial} translations. The
energy and momentum are invariant under coordinate transformations
not involving Lorentz rotations at spatial infinity, and behave as
a four-vector under the latter.  It is also possible to set up the
analysis of gravitational radiation in a fashion closely analogous
to electrodynamics by introducing a suitable definition of the
wave zone. In this region, gravitational waves propagate as free
radiation, independent of the strong field interior sources.  The
waves obey ordinary (flat-space) wave equations and consequently
satisfy superposition. The Poynting vector may also be defined
invariantly in the wave zone.  In contrast, the Newtonian-like
parts of the metric cannot be determined within the wave zone;
they depend strongly on the interior non-linearities. These points
are discussed in Section 5.

When the analysis is extended to include coupling of other systems
to the gravitational field (Section 6), the above definition of
energy may be used to discuss self-energy questions. In this way, the
static gravitation and electromagnetic self-masses of point
particle systems will be treated rigorously in Section 7. Here the
canonical formalism is essential in order that one may recognize a
pure particle (no wave) state.  The vanishing of the canonical
variables guarantees that there are no independent field
excitations contributing to the energy.  The total clothed mass of
a classical electron turns out to be finite, independent of its
bare mass and completely determined by its charge. Further, a
``neutral" particle (one coupled only to the gravitational field)
has a zero clothed mass, showing that the mass of a particle
arises entirely from its interactions with other fields.  The
physical origin of these finite results is discussed at the
beginning of Section 7 in terms of equivalence principle
considerations.  The self-stress ${\fT}^{ij}$ of the electron
vanishes, showing that the particle is stable, its repulsive
electrostatic self-forces being precisely cancelled by
gravitational attraction without any {\it ad hoc} compensation
being required.  Thus, a completely consistent classical point
charge exists when gravitation is included. These rigorous results
are in contrast to the higher and higher infinities that would
arise in a perturbation analysis of the same problem.

Whether gravitational effects will maintain the finiteness of
self-energies in quantum theory (and if so, whether the effective
cutoff will be appropriate to produce reasonable values) is at
present an open question.  In the final section (8), some
speculative remarks are made on this problem.  Since a complete
set of P.B.'s has been obtained classically in Section 4, it is
formally possible to quantize by the usual prescription of
relating them to quantum commutators.  However, the non-linear
nature of the theory may necessitate a more subtle transition to
the quantum domain. Section 8 also discusses some of these
questions.

\noindent{\large\bf 2. Classical Dynamics Background}

\renewcommand{\thefootnote}{\arabic{footnote}}
\setcounter{footnote}{0}

{\bf 2.1. {\it Action principle for Hamilton's equations}}. As was
mentioned in Section 1, general relativity is a theory in
``already parameterized form." We begin, therefore, with a brief
analysis of the relevant properties of the parameter formalism
(see also Lanczos, 1949). For simplicity, we deal with a system of
a finite number $M$ of degrees of freedom. Its action may be
written as
\renewcommand{\theequation}{2.\arabic{equation}}
\setcounter{equation}{0}
 \be%(2.1)
I = \int^{t{_1}}_{t{_2}} dt \: L = \int^{t{_1}}_{t{_2}} dt \:
\left( \sum^M_{i=1} p_i\dot q_i - H(p,q) \right)
 \ee
where $\dot q \equiv dq/dt$ and the Lagrangian has been expressed
in a form linear in the time derivatives. (This will be referred
to as the first-order form since independent variation of $p_i$
and $q_i$ gives rise to the first-order equations of motion.) The
maximal information obtainable from the action arises when not
only $p_i$ and $q_i$  are varied independently, but $t$ is also
varied and endpoint variations are allowed.  Postulating that the
total $\d I$ is a function only of the endpoints $[\d I = G(t_1) -
G(t_2)]$ leads to: (1) the usual Hamilton equations of motion for
$p_i$ and $q_i$; (2) conservation of energy $(d H/dt = 0)$; and
(3) the generating function
\be%2.2
G(t) = \sum_i p_i \, \d q_i - H \, \d t
 \ee
Here $\d q_i = \d_0 q_i + \dot q_i \, \d t$ where $\d_0 q_i$
denotes the independent (``intrinsic") variation of $q_i$. The
generating function can easily be seen to be the conventional
generator of canonical transformations. Thus $G_q = \sum_i p_i\,
\d q_i$  generates changes $q_i \rightarrow q_i + \d q_i \, , \;
p_i \rightarrow p_i$ while $G_t = - H \, \d t$ generates the
translation in time. That is, for $G_q$ one has $[q_j , G_q ] =
\sum_i [q_j,p_i]\d q_i =\d q_j$, where [$A,B$] means the Poisson
bracket (P.B.), and for $G_t$ one has $[q_i , G_t] = - [q_i,H]\d t
= -\dot q_i \, \d t$ by the P.B. form of the equations of motion.
The above elementary discussion may be inverted to show that, for
the action of (2.1), if every variable occurring in $H$ is also
found in the $p\dot q$ term, then the theory is in canonical form
and $p_i$ and $q_i$ obey the conventional P.B. relations. This is
the classical equivalent of the Schwinger action principle
(Schwinger, 1951, 1953).

{\bf 2.2. {\it The action in parameterized form}}. The motion of
the system (2.1) is described in terms of one independent variable
$t$ (the ``coordinate").  The action may be cast, as is well
known, into parameterized form, in which the time is regarded as a
function $q_{M+1}$ of an arbitrary parameter $\t$:
$$
I = \int^{\t_1}_{\t_2} d\t \, L_\t \equiv \int^{\t_1}_{\t_2} d\t
\left[ \sum^{M+1}_{i=1} p_i q^\prime_i \right] \; .
 $$
Here, $q^\prime \equiv dq/d\t$, and the constraint equation
$p_{M+1} + H(p, q) =0$ holds.  One may equally well replace this
constraint by an additional term in the action:
\be%2.3
I = \int^{\t_1}_{\t_2} d\t \left[ \sum^{M+1}_{i=1} p_i q^\prime_i
- NR \right]
 \ee
where $N(\t )$ is a Lagrange multiplier. Its variation yields the
constraint equation $R(p_{M+1}, p, q) = 0$, which may be any
equation with the solution (occurring as a simple root) $p_{M+1} =
-H$. The theory as cast into form (2.3) is now generally covariant
with respect to arbitrary coordinate transformations $ \bar \t =
\bar \t (\t )$, bearing in mind that $N$ transforms as $dq/d\t$.
The price of achieving this general covariance has been not only
the introduction of the $(M + 1)$st degree of freedom, but, more
important, the loss of canonical form, due to the appearance of
the Lagrange multiplier $N$ in the ``Hamiltonian," $H^\prime
\equiv NR$. ($N$ occurs in $H^\prime$ but not in $\sum^{M+1}_{i=1}
p_iq^\prime_i$.) A further striking feature which is due to the
general covariance of this formulation is that the ``Hamiltonian"
$H^\prime$ vanishes by virtue of the constraint equation. This is
not surprising, since the motion of any particular variable $F(p,
q)$ with respect to $\t$ is arbitrary, {\it i.e.}, $F^\prime$ may
be given any value by suitable recalibration $\t \rightarrow
\bar\t$.

{\bf 2.3. {\it Reduction of parameterized action to Hamiltonian
form---intrinsic coordinates.}} As we shall see, the Lagrangian of
general relativity may be written in precisely the form of (2.3).
We will, therefore, be faced with the problem of reducing an
action of the type (2.3) to canonical form (2.1). The general
procedure consists essentially in reversing the steps that led to
(2.3). If one simply inserts the solution, $p_{M+1} = -H$, of the
constraint equation into (2.3), one obtains
\be%2.4
I = \int d\t \left[ \sum^M_{i=1} p_i q^\prime_i -
H(p,q)q^\prime_{M+1} \right] \; .
 \ee
All reference to the arbitrary parameter $\t$ disappears when $I$
is rewritten as
\be%2.5
I = \int dq_{M+1}  \left[ \sum^M_{i=1} p_i (dq_i/dq_{M+1})-H
\right]
 \ee
which is identical to (2.1)with the notational change $q_{M+1}
\rightarrow t$. Equation (2.5) exhibits the role of the variable
$q_{M+1}$ as an ``intrinsic coordinate." By this is meant the
following. The equation of motion for $q_{M+1}$ is
$q^\prime_{M+1} = N(\pa R/\pa q_{M+1})$ from (2.3). Also, none of
the dynamical equations determine $N$ as a function of $\t$. Thus
$N$ and hence $q_{M+1}$, are left arbitrary by the dynamics
(though, of course, a choice of  $q_{M+1}$  as a function of $\t$
fixes $N$). One is therefore free to choose $q_{M+1} (\t )$ to be
any desired function and use this function as the new independent
variable (parameter):  $q_i = q_i( q_{M+1})$, $p_i = p_i(
q_{M+1})$, $i = I \ldots M$. The action of (2.5), and hence the
relations between $q_i$, $p_i$, and  $q_{M+1}$ are now independent
of $\t$. They are manifestly invariant under the general
``coordinate transformation" $\bar\t = \bar\t (\t )$ (for the
simple reason that $\t $ itself no longer appears). The choice of
$q_{M+1}$ as the independent variable thus yields a manifestly
$\t$-invariant formulation and gives an ``intrinsic" specification
of the dynamics.  This is in contrast to the original one in which
the trajectories of $q_1 \ldots q_{M+1}$  are given in terms of
some arbitrary variable $\t$ (which is extraneous to the system).

In practice, we shall arrive at the intrinsic form (2.5) from
(2.4) in an alternate way.  Since the relation between $q_{M+1}$
and $\t$ is undetermined, we are free to specify it explicitly,
{\it i.e.}, impose a ``coordinate condition." If, in particular,
this relation is chosen to be $q_{M+1}=  \t$ (a condition which
also determines $N$), the action (2.4) then reduces (2.5) with the
notational change $q_{M+1} \rightarrow \t$; the non-vanishing
Hamiltonian only arises as a result of this process. [Of course,
other coordinate conditions might have been chosen. These would
correspond to using a variable other than $q_{M+1}$ as the
intrinsic coordinate in the previous discussion.]

This simple analysis has shown that the way to reduce a
parameterized action to canonical form is to insert the solution
of the constraint equations and to impose coordinate conditions.
Further, the imposition of coordinate conditions is equivalent to
the introduction of intrinsic coordinates.

In field theory it will prove more informative to carry out this
analysis in the generator. We exhibit here the procedure in the
particle case: The generator associated with the action of (2.3)
is
\be%2.6
G = \sum^{M+1}_{i=1} p_i \, \d q_i - NR \, \d\t
 \ee
Upon inserting constraints, the generator reduces to
\be%2.7
G = \sum^M_{i=1} p_i \, \d q_i - H \, \d q_{M+1} \; .
 \ee
Imposing the coordinate condition $q_{M+1} = t$ then yields (2.2)
From {\it this} form, one can immediately recognize the $M$ pairs
of canonical variables and the non-vanishing Hamiltonian of the
theory.

One can, of course, perform the above analysis for a parameterized
field theory as well.  Here the coordinates appear as four new
field variables  $q^{M+\m} = x^\m (\t^\a )$, and there are four
extra momenta  $p_{M+\m}(\t^\a )$ conjugate to them. Four
constraint equations are required to relate these momenta to the
Hamiltonian density and the field momentum density, and
correspondingly, there are four Lagrange multipliers $N_\m (\t ^\a
)$ for a field.  An example in which the scalar meson field is
parameterized may be found in III.

\noindent{\large\bf 3. First-Order Form of the Gravitational
Field}
\renewcommand{\theequation}{3.\arabic{equation}}
\setcounter{equation}{0}

{\bf 3.1. {\it The Einstein action in first-order (Palatini)
form.}} The usual action integral for general
relativity\footnote{We use units such that 16$\pi\g c^{-4} = 1 =
c$, where $\g$ is the Newtonian gravitational constant; electric
charge is in rationalized units. Latin indices run from 1 to 3,
Greek from 0 to 3, and $x^0 = t$. Derivatives are denoted by a
comma or the symbol $\pa_\m$.}
\be%3.1
I = \int d^4x \, {\cal L} = \int d^4x\, \sqrt{-g}\: R
 \ee
yields the Einstein field equations when one considers variations
in the metric ({\it e.g.}, $g_\mn$ or the density  $\fg^\mn =
\sqrt{-g} \: g^\mn$). These Lagrange equations of motion are then
second-order differential equations.  It is our aim to obtain a
canonical form for these equations, that is, to put them in the
form  $\dot q = \pa H/\pa p,\; \dot p = -\pa H/\pa q$.  As a
preliminary step, we will restate the Lagrangian so that the
equations of motion have two of the properties of canonical
equations: (1) they are first-order equations; and (2) they are
solved explicitly for the time derivatives.  The second property
will be obtained by a 3 + 1 dimensional breakup of the original
four-dimensional quantities, as will be discussed below. The first
property is insured by using a Lagrangian linear in first
derivatives.  In relativity, this is called the Palatini
Lagrangian, and consists in regarding the Christoffel symbols
$\G_\m\,^\a\,_\n$ as independent quantities in the variational
principle (see, for example, Schr\"odinger, 1950). Thus, one may
rewrite (3.1) as
 \be%3.2
  I = \int d^4 x \, {\fg}^\mn R_\mn (\G )
 \ee
 where
\be%3.3
R_\mn (\G ) \equiv \G_\m\,^\a\,_{\n,\a} - \G_\m\,^\a\,_{\a,\n} +
\G_\m\,^\a\,_\n \G_\a\,^\b\,_\b - \G_\m\,^\a\,_\b \G_\n\,^\b\,_\a
\; .
 \ee
Note that these covariant components $R_\mn$ of the Ricci tensor
do not involve the metric but only the affinity $\G_\m\,^\a\,_\n$.
Thus, by varying $g^\mn$, one obtains directly the Einstein
field equations
 $$%3.4a
G_\mn \equiv R_\mn - \textstyle{\frac{1}{2}} \: g_\mn \, R = 0 \;
. \eqno{(3.4a)}
 $$
These equations no longer express the full content of the theory,
since the relation between the now independent quantities
$\G_\m\,^\a\,_\n$ and $g_\mn$ is still required.  This is obtained
as a field equation by varying $\G_\m\,^\a\,_\n$. One then finds
 $$%3.4b
\fg^\mn\,\!_{;\a} \equiv \fg^\mn\,\!_{,\a} + \fg^{\m\b}\G_\b\,^\n\,_\a +
\fg^{\n\b}\G_\b\,^\m\,_\a - \fg^\mn\G_\b\,^\b\,_\a = 0
 \eqno{(3.4b)}
 $$
 which, as is well known, can be solved for $\G_\m\,^\a\,_\n$ to give
the usual relation  $\G_\m\,^\a\,_\n =
\left\{_\m\,\!\!^\a\,\!\!_\n\! \right\} \equiv \frac{1}{2} \,
g^\ab (g_{\m\b ,\n} + g_{\n\b ,\m} - g_{\mn ,\b} )$.

\renewcommand{\theequation}{3.\arabic{equation}}
\setcounter{equation}{4}

The Palatini formulation of general relativity has a direct analog
in Maxwell theory, the affinity corresponding to the field
strength $F_\mn$ and the metric to the vector potential $A_\m$.
Here the Lagrangian is
 \be%3.5
{\cal L} = A_{\m ,\n} F^\mn + \textstyle{\frac{1}{4}} \, F_\mn
F^\mn
 \ee
with $A_\m$ and $F_\mn$ to be independently varied.  The field
equations then become
 $$%3.6a
F^\mn\,_{,\n} = 0 \eqno{(3.6a)}
 $$
  and
  $$%3.6b
A_{\n ,\m} - A_{\m ,\n} = F_\mn \eqno{(3.6b)}
 $$
which correspond to (3.4$a$) and (3.4$b$).

The next step in achieving canonical form is to single out time
derivatives by introducing three-dimensional notation. For the
Maxwell field, we thus define
 $$%3.7a
 E^i \equiv F^{0i} \; . \eqno{(3.7a)}
 $$
Since the canonical form requires equations of first order in time
derivatives (but not in space derivatives), we may use the
equation $F_{ij} = A_{j,i} - A_{i,j}$ to eliminate $F_{ij}$. In
terms of the abbreviation
 $$%3.7b
 B^i \equiv  \textstyle{\frac{1}{2}} \: \e^{ijk} (A_{k,j} -
 A_{j,k}) \eqno{(3.7b)}
  $$
 the Lagrangian reads
 \renewcommand{\theequation}{3.\arabic{equation}}
\setcounter{equation}{7}
\be
 {\cal L} = - E^i \: \pa_0 A_i - \textstyle{\frac{1}{2}} \:
 (B^iB^i + E^iE^i) - A_0 E^i\,_{,i} \; . \ee
At this stage, the Maxwell equations are obtained by varying
${\cal L}$ with respect to the independent quantities $E^i,\;
A_i$, and $A_0$.

{\bf 3-2. {\it Three-plus-one dimensional decomposition of the
Einstein field.}}  The three-dimensional quantities appropriate
for the Einstein field are (as will be discussed in detail later)
 $$%3.9a
g_{ij} \equiv \; ^4g_{ij} \; , \hspace{.4in} N \equiv
(-^4g^{00})^{-1/2} \; , \hspace{.4in} N_i \equiv \;^4g_{0i}
\eqno{(3.9a)}
  $$
 $$%3.9b
\pi^{ij} \equiv \sqrt{-^4g} \: (^4\G_p\,^0\,_q -
g_{pq}\,^4\G_r\,^0\,_s g^{rs}) g^{ip}g^{jq}\; .
 \eqno{(3.9b)}
  $$
Here and subsequently we mark every four-dimensional quantity with
the prefix $^4$, so that all unmarked quantities are understood as
three-dimensional.  In particular, $g^{ij}$ in (3.9b) is the
reciprocal matrix to  $g_{ij}$  . The full metric $^4g_\mn$ and
$^4g^\mn$ may, with (3.9a), be written
 \renewcommand{\theequation}{3.\arabic{equation}}
\setcounter{equation}{9}
 \be%3.10
 ^4g_{00} = - (N^2 - N_iN^i)
 \ee
  where $N^i = g^{ij}N_j$, and
 $$%3.11a
 ^4g^{0i} = N^i/N^2 \; , \hspace{.4in} ^4g^{00} = -1/N^2
 \eqno{(3.11a)}
  $$
 $$%3.11b
 ^4g^{ij} = g^{ij} - (N^iN^j/N^2) \; .
 \eqno{(3.11b)}
  $$
 One further useful relation is
 \renewcommand{\theequation}{3.\arabic{equation}}
\setcounter{equation}{11}
 \be%3.12
\sqrt{-^4g} = N\sqrt{g} \; .
 \ee

In terms of the basic quantities of (3.9), the Lagrangian of
general relativity becomes
 \bea%3.13
\lefteqn{{\cal L} = \sqrt{-^4g} \; ^4R = -g_{ij} \, \pa_t\pi^{ij}
- NR^0 - N_iR^i}~~~~~~~~~~~~~~~~~~ \nonumber \\
&& - 2 (\pi^{ij} N_j - \textstyle{\frac{1}{2}} \pi N^i + N^{|i}
\sqrt{g} )_{,i}
 \eea
where
 $$%3.14a
R^0 \equiv - \sqrt{g} \: [^3R + g^{-1} (\textstyle{\frac{1}{2}}\:
\pi^2 - \pi^{ij}\pi_{ij} )]
 \eqno{(3.14a)}
  $$
 $$%3.14b
R^i \equiv - 2  \pi^{ij}\,_{|j}\; . ~~~~~~~~~~~~~~~~~~~~~~~~~~~~
 \eqno{(3.14b)}
  $$
The quantity $^3R$ is the curvature scalar formed from the spatial
metric $g_{ij}, \; _|$ indicates the covariant derivative using
this metric, and spatial indices are raised and lowered using
$g^{ij}$ and $g_{ij}$. (Similarly, $\pi \equiv \pi^i\,_i$.) As in
the electromagnetic example, we have allowed second-order space
derivatives to appear by eliminating such quantities as
$\G_i\,^k\,_j$ in terms of $g_{ij,k}$.

One may verify directly that the first-order Lagrangian (3.13)
correctly gives rise to the Einstein equations.  One obtains
 $$%3.15a
 \pa_t g_{ij} = 2Ng^{-1/2}(\pi_{ij} - \textstyle{\frac{1}{2}}\:
 g_{ij}\pi ) + N_{i|j} + N_{j|i}~~~~~~~~~~~~~~~~~~~~~~~~~~~~~~ \eqno{(3.15a)}
 $$
$$%3.15b
\pa_t \pi^{ij}  =  - \; N\sqrt{g} \: (^3\!R_{ij} -
\textstyle{\frac{1}{2}}\:  g^{ij}\: ^3\!R) +
\textstyle{\frac{1}{2}}\: Ng^{-1/2} g^{ij} (\pi^{mn}\pi_{mn}
-\textstyle{\frac{1}{2}}\pi^2 )~~~
$$
$$
\hspace*{.3in}- \; 2Ng^{-1/2} (\pi^{im}\pi_m\,^j
-\textstyle{\frac{1}{2}}\pi\pi^{ij}) + \sqrt{g} \: (N^{|ij} -
g^{ij} \, N^{|m}\,_{|m}) \eqno{(3.15b)}
$$
$$
 + \; (\pi^{ij} N^m)_{|m} - N^i_{|m}\pi^{mj} -
N^j\,_{|m}\pi^{mi}~~~~~~~~~~~~~~~~~~
$$
$$%3.15c
R^\m (g_{ij},\pi^{ij}) = 0 \; . \eqno{(3.15c)}
$$
Equation (3.15a), which results from varying $\pi^{ij}$, would be
viewed as the defining equation for  $\pi^{ij}$  in a second-order
formalism. Variation of $N$ and $N_i$ yields equations (3.15c),
which are the $^4\!G^0\,_\m \equiv \;^4\!R^0\,_\m -
\frac{1}{2}\:\d^0_\m \:^4\!R = 0$ equations, while equations
(3.15b) are linear combinations of these equations and the
remaining six Einstein equations $(^4G_{ij}=0)$.

{\bf 3.3. {\it Geometrical interpretation of dynamical
variables.}} Before proceeding with the reduction to canonical
form, it is enlightening to examine, from a geometrical point of
view, our specific choices (3.9) of three-dimensional variables.
Geometrically, their form is governed by the requirement that the
basic variables be three-covariant under all coordinate
transformations which leave the $t$=const surfaces unchanged. Any
quantities which have this property can be defined entirely within
the surface (this is clearly appropriate for the 3+1 dimensional
breakup).  One fundamental four-dimensional object which is
clearly also three-dimensional is a curve $x^\m (\l)$ which lies
entirely within the 3-surface, {\it i.e.}, $x^0 (\l)$  = const.
The vector $v^\m \equiv dx^\m /d\l$ tangent to this curve is
therefore also three-dimensional. The restriction that the curve
lie in the surface $t$=const is then $v^0 = 0$, and conversely
{\it any} vector $V^\m$, with $V^0 = 0$ is tangent to some curve
in the surface. Three such independent vectors are $V^\m_{(i)} =
\d^\m_i$. Given any covariant tensor $A_{\m\ldots\n}$, its
projection onto the surface is then $V^\m_{(i)}\ldots
V^\n_{(j)}A_{\m\ldots\n} = A_{i\ldots j}$. Thus, the {\it
covariant} spatial components of any four-tensor form a
three-tensor which depends only on the surface\footnote{Of course,
any three-dimensional operations (such as contraction of indices
with the three-metric) on a three-tensor yield quantities defined
on the three-surface.} (in contrast to the contravariant spatial
components which are scalar products with gradients rather than
tangents, and hence depend also on the choice of spatial
coordinates in the immediate neighborhood of the surface). This
accounts for the choice of $g_{ij}$, rather than $^4\!g^{ij}$. In
contrast, $N$ and $N_i$ do not have the desired invariance and, in
fact, by choosing coordinates such that the $x^i$ = const lines
are normal to the surface, one obtains $N_i = 0$. (If $x^0$ is
arranged to measure proper time along these lines, one has also $N
= 1$.) By the same argument, one can see that $A_i$ and $F_{ij}$
are appropriate three-dimensional quantities in a general
relativistic discussion of the Maxwell field.

The quantity which plays the role of a momentum is more difficult
to define within the surface, since it refers to motion in time
leading out of the original $t =$ const surface.  Such a quantity
is, however, provided by the second fundamental form $K_{ij}$
(see, for example, Eisenhart, 1949), which gives the radii of
curvature of the $t =$ const surface as measured in the
surrounding four-space. These ``extrinsic curvatures" describe how
the normals to the surface converge or diverge, and hence
determine the geometry of a parallel surface at an infinitesimally
later time. Since $K_{ij}$ describes a geometrical property of the
$t =$ const surface, as imbedded in four-space, it again does not
depend on the choice of coordinates away from the surface. This
may also be seen from a standard definition, $K_{ij} = -
n_{(i;j)}$, which expresses $K_{ij}$ as the covariant spatial part
of the tensor $n_{(\m ;\n )}$  (the four-dimensional covariant
derivative of the unit normal,  $n_\m = - N\d^0_\m$ to the
surface).\footnote{Thus one has $K_{ij} = - n_{(i;j)} = -
n_{(i,j)}
 + n_\m\G_i\,^\m\,_j = N\G_i\,^0\,_j$.}
For convenience in ultimately reaching canonical form, we have
chosen, instead of $K_{ij}$, the closely related variable
$\pi^{ij} = - \sqrt{g} \: (K^{ij} - g^{ij} K)$. Thus, the
geometrical analysis defines $g_{ij}$ and $g_{ij}$  as suitable
quantities, unaffected by the choice of coordinates later in time,
while $N$ and $N_i$ describe how the coordinate system will be
continued off the $t =$ const surface.

{\bf 3.4. {\it Initial value problem and dynamical structure of
field equations.}} Returning to the field equations (3.15), we may
now analyze them from the point of view of the initial value
problem. If one specifies $g_{ij}, \; \pi^{ij}$ and $N,\; N_i$
initially, it is clear that the equations uniquely determine
$g_{ij}$ and $\pi^{ij}$ at a later time, while $N$ and $N_i$
remain undetermined then. Since the latter merely express the
continuation of the coordinates, the intrinsic
(coordinate-independent) geometry of space-time is determined
uniquely by an initial choice of $g_{ij}$ and $\pi^{ij}$.  This
choice is restricted, however, by the four constraint equations
(3.15c) which relate these twelve variables at the initial time.
Subject to these compatibility conditions, then, the ($g_{ij}, \;
\pi^{ij}$) constitute a complete set of Cauchy data for the
theory.

The maintenance in time of these constraints is guaranteed by the
Bianchi identities $(^4\!G_\m\!\,^0\!\,_{;\n}\equiv 0)$.  Hence
 \renewcommand{\theequation}{3.\arabic{equation}}
\setcounter{equation}{15}
 \be%3.16
 ^4\!G_\m\!\,^0\!\,_{,0} = \;^4\!G_\m\!\,^i\,_{;i} - \;^4\!G_\m\!\,^\n \:
^4\!\G_\n\,^0\,_0 + \;^4\!G_\n\,^0 \:^4\!\G_\m\!\,^\n\!\,_0 \; .
 \ee
Thus, by virtue of the dynamic equations  $^4\!G_{ij}=0$ (and
consequently of their spatial derivatives) at $t = 0$, the
constraints  $^4\!G_\m\,^0=0$  hold at all times if they hold
initially.

In electrodynamics, the constraint equation corresponding to
(3.15c) is obtained by varying $A_0$ in (3.8), and is
$F^{0i}\!\!\,_{,i} \equiv \mbox{\boldmath $\nabla\cdot E$} = 0$.
The identity ensuring its maintenance in time is just
$F^\mn\!\!\,_{,\mn} \equiv 0$, which may be rewritten as
$(F^{0i}\,\!\!_{,i})_{,0} = -(F^{i\n}\!\!\,_{,\n})_{,i}$. The
right-hand side then vanishes by the dynamic equations
$F^{i\n}\!\!\,_{,\n} =0$.

While the twelve variables  ($g_{ij}, \; \pi^{ij}$)  constitute a
complete set of Cauchy data, they do not provide a minimal set
(which the canonical formalism will eventually give to be two
pairs, corresponding to two degrees of freedom).  We may now count
the number of minimal variables.  Of the twelve  $g_{ij}, \;
\pi^{ij}$, we may eliminate four by using the constraint equations
(3.15c). There will correspondingly be four ``Bianchi" identities
among the twelve equations of motion (3.15a) and (3.15b). As we
have seen, $N$ and $N_i$ determine the continuation of the
coordinate system without affecting the intrinsic geometry ({\it
i.e.}, the physics of the field). For every choice of $N$ and
$N_i$ as functions of the remaining eight Cauchy data (which
represents a choice of coordinate frame), there will result four
equations stating that the time derivatives of four of the
remaining eight ($g, \;\pi$) variables vanish. [More precisely, a
choice of coordinate frame is made by specifying four functions
$q^\m$ of  ($g_{ij}, \; \pi^{ij}$) as the coordinates $x^\m$.  The
equations for $\pa_tq^\m \; (=\d^\m_0 )$ then determine $N$ and
$N_i$.] Thus, after these coordinate conditions are imposed, we
are left with four dynamic equations of the form $\pa_t u_a =
f_a(u) \; (a = 1, 2, 3, 4)$. These equations govern the motion of
a system of two degrees of freedom. This is to be expected, since
the linearized gravitational field is a massless spin two field,
and the self-interaction of the full theory should not alter such
kinematical features as the number of degrees of freedom.

{\bf 3.5. {\it The gravitational field as an already parameterized
system.}} To conclude this section, we point out the
characteristic properties of the Einstein Lagrangian of (3.13). We
reproduce it here with a divergence\footnote{A divergence in a
theory with constraints cannot necessarily be discarded, since
such a term may cease to be a divergence upon elimination of
constraints.  In IIIA, it is shown that the divergences neglected
in this work are indeed to be discarded.} and a total time
derivative\footnote{The addition of a total time derivative in a
Lagrangian does not, of course, alter the equations of motion and
corresponds to a canonical transformation in the generator. The
properties of canonical transformations in general relativity are
discussed in IVA.} discarded:
 \be%3.17
{\cal L} = \pi^{ij} \: \pa_t g_{ij} - NR^0 - N_iR^i \;
,\hspace{.4in} R^\m = R^\m (g_{ij}, \: \pi^{ij}) \; .
 \ee
Equation (3.17) is thus precisely in the form of a parameterized
theory's Lagrangian as in (2.3). This form just expresses the
invariance of the theory with respect to transformations of the
four coordinates $x^\m$ and hence the $x^\m$ are parameters in
exactly the same sense that $\t$ was in the particle case.  That
the $N$ and $N_i$ are truly Lagrange multipliers follows from the
fact that they do not appear in the $pq^\prime$ ({\it i.e.},
$\pi^{ij} \: \pa_tg_{ij}$) part of ${\cal L}$. Their variation
yields the four constraint equations $R^\m$, = 0. The
``Hamiltonian" ${\cal H}^\prime \equiv NR^0 + N_iR^i$ vanishes due
to the constraints. The true non-vanishing Hamiltonian of the
theory will arise only after the constraint variables have been
eliminated and coordinate conditions chosen. The analysis leading
to the canonical form is carried out in the next section.

\noindent{\large\bf 4. Canonical Form for General Relativity}

{\bf 4.1. {\it Analysis of generating functions.}} We are now in a
position to cast the general theory into canonical form. The
geometrical considerations of Section 3 were useful in obtaining
the Lagrangian in the form (3.17), which, in the light of Section
2, we recognized as the Lagrangian of a parameterized field theory
corresponding to (2.3).

The reduction of (3.17) to the canonical form analogous to (2.1)
requires an identification of the four extra momenta to be
eliminated by the constraint equations (3.15c). To this end we
consider the generator arising from (3.17):
 \renewcommand{\theequation}{4.\arabic{equation}}
\setcounter{equation}{0}
 \be%4.1
 G = \int d^3x \: [\pi^{ij} \: \d g_{ij} + T^0\,\!_\m\,\!^\prime
 \: \d x^\m ]\; .
 \ee
The $T^0\,\!_\m\,\!^\prime \: \d x^\m$ term comes from the
independent coordinate variations. However,
$T^0\,\!_\m\,\!^\prime$ vanishes as a consequence of the
constraint equations.\footnote{More precisely
$T^0\,\!_i\,\!^\prime$ reduces to an irrelevant divergence.} For
example, $T^0\,\!_0\,\!^\prime = - NR^0 -N_iR^i = 0$. When the
constraints are inserted, $G$ reduces to
 \be%4.2
G = \int d^3x \: \pi^{ij} \: \d g_{ij}
 \ee
[corresponding to (2.7)] where four of the twelve ($g_{ij}, \;
\pi^{ij}$) are understood to have been expressed in terms of the
rest by solving $R^\m = 0$ for them. This elimination exhausts the
content of the constraint equations. Finally, as in the particle
case, coordinate conditions (now four in number) must be chosen
and this information inserted into (4.2), leaving one now with
only four dynamical variables ($\pi^A, \;\phi_A$). If, in fact,
the generator at this stage has the form
 \be%4.3
 G = \int d^3x \: \left[ \sum^2_{A=1} \pi^A \: \d\phi_A + 
       {\fT}^0\,\!_\m (\pi^A, \:\phi_A ) \d x^\m \right]
 \ee
then the theory is clearly in {\it canonical} form with $\pi^A$
and $\phi_A$ as canonical variables and $\int {\fT}^0\,\!_\m \:
\d x^\m$ the generator of translations $\d x^\m$.  In (4.3),
${\fT}^0\,\!_\m [\pi^A, \:\phi_A]$ has arisen in the
elimination of the extra momenta $p_{M+\m}$, by solving the
constraint equations, and the $x^\m$, now represent the four
variables chosen as coordinates $q_{M+\m}$.

For the Maxwell field, the generator corresponding to (4.2) is, by
(3.8),
$$
G = \int d^3x [-E^i \: \d A_i - \textstyle{\frac{1}{2}} (E^iE^i +
B^iB^i) \d t + ({\bf E} \mbox{\boldmath$\times$} {\bf B})\cdot \d
{\bf r}]
$$
with the constraint $E^i\,\!_{,i} = 0$ understood to have been
eliminated. The solution of the constraint equation is, of course,
well known in terms of the orthogonal decomposition $E^i = E^{iT}
+E^{iL}\; (\nabla  \mbox{\boldmath$\cdot$}{\bf E}^T \equiv 0
\equiv \nabla \mbox{\boldmath$\times$} {\bf E}^L)$. The
longitudinal component $E^{iL}$ of the electric field vanishes,
and therefore the canonical form is reached when one inserts this
information into $G$:
 $$
G = \int d^3x [-E^{iT} \: \d A_i\,\!^T - \textstyle{\frac{1}{2}}
(E^{iT}E^{iT} + B^iB^i) \d_t + ({\bf E}^T \mbox{\boldmath$\times$}
{\bf B})\cdot \d {\bf r}] \; .
$$
Note that $A_i\,\!^L$ has automatically disappeared from the
kinetic term (by orthogonality), and never existed in ${\bf B}
\equiv \nabla  \mbox{\boldmath$\times$} {\bf A} = \nabla
\mbox{\boldmath$\times$} {\bf A}^T$. Thus, the two canonically
conjugate pairs of Maxwell field variables are $-E^{iT}$ and
$A_i\,\!^T$.

{\bf 4.2. {\it Analysis of constraint equations in linearized
theory-orthogonal decomposition of the metric.}}  In order to
achieve the form (4.3) for relativity, it is useful to be guided
by the linearized theory. Here one must treat the constraint
equations to second order since our general formalism shows that
the Hamiltonian arises from them. Through quadratic terms,
equations (3.15c) may be written in the form
$$%4.4a
g_{ij,ij} -g_{ii,jj} = {\cal P}_2\,\!^0[g_{ij},\pi^{ij}]~~~~~~~~
\eqno{(4.4a)}
$$
$$%4.4b
-2\pi^{ij}\,\!_{,j} = \; {\cal P}_2\,\!^i[g_{ij},\pi^{ij}]
\eqno{(4.4b)}
$$
where ${\cal P}_2\,\!^0$ and ${\cal P}_2\,\!^i$ are purely
quadratic functions of $g_{ij}$ and $\pi^{ij}$. These equations
determine one component of $g_{ij}$ and three components of
$\pi^{ij}$ in terms of the rest.  The content of equations (4.4)
can be seen more easily if one makes the following linear
orthogonal decomposition on  $g_{ij}$ and $\pi^{ij}$. For any
symmetric array $f_{ij} = f_{ji}$ one has
 \renewcommand{\theequation}{4.\arabic{equation}}
\setcounter{equation}{4}
 \be%4.5
f_{ij} = f_{ij}\,\!^{TT} + f_{ij}\,\!^T + (f_{i,j} + f_{j,i})
 \ee
where each of the quantities on the right-hand side can be
expressed uniquely as a linear functional of $f_{ij}$.  The
quantities $f_{ij}\,\!^{TT}$ are the two transverse traceless
components of $f_{ij}(f_{ij}\,\!^{TT}\,\!_{,j} \equiv 0,
\;f_{ii}\,\!^{TT}  \equiv 0)$. The trace of the transverse part of
$f_{ij}$, {\it i.e.}, $f^T$, uniquely defines $f_{ij}\,\!^T$
according to
 \be%4.6
f_{ij}\,\!^T \equiv \textstyle{\frac{1}{2}} [\d_{ij} f^T -
(1/\nabla^2)f^T\,\!_{,ij}]
 \ee
(and clearly, $f_{ij}\,\!^T\,\!_{,j} = 0, \; f_{ii}\,\!^T = f^T$.
The operator $1/\nabla^2$ is the inverse of the flat space
Laplacian, with appropriate boundary conditions. The longitudinal
parts of $f_{ij}$ reside in the remaining part, $f_{i,j} +
f_{j,i}$.  Decomposing $f_i$ into its transverse and longitudinal
(curl-less) parts, one has  $f_i = f_i\,\!^T +
\textstyle{\frac{1}{2}} f^L\,\!_{,i} \; (f_i\,\!^T\,\!_{,i} \equiv
0)$. The remainder then becomes $f_i\,\!^T\,\!_{,i} +
f_j\,\!^T\,\!_{,i} + f^L\,\!_{,ij}$. One may express $f_i, \; f^T$
and $f_{ij}\,\!^{TT}$ in terms of $f_{ij}$  by
 $$%4.7a
f_i = (1/\nabla^2) [f_{ij,j} - \textstyle{\frac{1}{2}}\,
(1/\nabla^2)f_{kj,kji}]~~~~~~~~ \eqno{(4.7a)}
$$
 $$%4.7b
f^T = f_{ii} - (1/\nabla^2)f_{ij,ij}~~~~~~~~~~~~~~~~~~~~~~~~~~
\eqno{(4.7b)}
$$
 $$%4.7c
f_{ij}\,\!^{TT} = f_{ij} - f_{ij}\,\!^T [f_{mn}] - \{
f_{i,j}[f_{mn}]  +  f_{j,i}[f_{mn}] \} .\eqno{(4.7c)}
$$
The foregoing orthogonal decomposition for a symmetric tensor is
just the extension of the usual decomposition of a vector into
longitudinal and transverse parts employed in electromagnetic
theory.

Returning to (4.4), one has
 $$%4.8a
~~~~~~~~~~~~ - \nabla^2 g^T = {\cal P}_2\,\!^0 \eqno{(4.8a)}
 $$
 $$%4.8b
- 2\nabla^2 (\pi^{iT} + \pi^L\,\!_{,i}) = {\cal P}_2\,\!^i \; .
\eqno{(4.8b)}
$$
To first order, one sees that $g^T$ and $\pi^i$
vanish.\footnote{We use boundary conditions such that  $g^T$ and
$\pi^i$ vanish asymptotically.  Note also that $\pi^{iT}$  and
$\pi^L$ are independently determined by (4.8b).} These structures
begin, therefore, at second order where $g^T = - (1/\nabla^2){\cal
H}_{{\rm lin}}$ and  $-2 (\pi^{iT} + \pi^L\,\!_{,i}) =
(1/\nabla^2){\fT}^{0i}_{{\rm lin}}$. Here ${\cal H}_{{\rm
lin}}$, and ${\fT}^{0i}_{{\rm lin}}$ are obtained from ${\cal
P}^0_2$ and  ${\cal P}^i_2$ by setting $g^T$ and $\pi^i$ equal to
zero there. They are just the linearized theory's Hamiltonian and
field momentum densities.\footnote{More precisely, ${\cal
H}_{{\rm lin}}$ differs from the linearized Hamiltonian density by
a divergence which vanishes when coordinate conditions (4.11) are
imposed.} The analysis has thus shown that the constraint
equations can be solved for $g^T$ and $\pi^i$  (in terms of the
remaining variables) as the four extra momenta. To see explicitly
that ${\cal H}_{{\rm lin}}$, and ${\fT}^{0i}_{{\rm lin}}$
generate the appropriate time and space translations, one must
return to the generator. Inserting the orthogonal decomposition
(4.5) for both $g_{ij}$ and $\pi^{ij}$  into (4.2), one obtains
 \renewcommand{\theequation}{4.\arabic{equation}}
\setcounter{equation}{8}
 \be%4.9
G = \int d^3x \: [\pi^{ijTT} \: \d g_{ij}\,\!^{TT} + \pi^{ijT} \:
\d g_{ij}\,\!^T + 2 (\pi^i\,\!_{,j} +\pi^j\,\!_{,i})\d g_{i,j}] \;
.
 \ee
The cross-terms in (4.9) have vanished due to the orthogonality of
the decomposition ({\it e.g.}, $\int d^3x$  $\pi^{ijTT}\d g_{i,j}
= - \int d^3x \: \pi^{ijTT}\,\!_{,j} \: \d g_i = 0$).  We have
also used here the fact that taking the variation of a quantity
does not alter its transverse or longitudinal character in such a
linear breakup and that the derivatives commute with the
variation. Equation (4.9) may be brought to the desired form by
further integration by parts and addition of a total
variation:\footnote{Addition of a total variation to a generator
corresponds to the addition of a total time derivative in the
Lagrangian, as stated earlier.}
 \be%4.10
G = \int d^3x \: \{ \pi^{ijTT} \: \d g_i\,\!^{TT} - (-\nabla^2
g^T) \d [-(1/2\nabla^2)\pi^T]
 + \; [-2\nabla^2 (\pi^{iT} + \pi^L\,\!_{,i})]\} \d g_i
 \ee

{\bf 4.3. {\it Imposition of coordinate conditions.}} Equation
(4.10) is now in the form of (2.7). The final step in reduction to
canonical form is to impose coordinate conditions. The structure
of (4.10)suggests that one choose as coordinate conditions
$$%4.11a
t= - (1/2\nabla^2)\pi^T\eqno{(4.11a)}
$$
$$%4.11b
x^i = g_i \; . ~~~~~~~~~~~~~~ \eqno{(4.11b)}
$$
Alternately, these coordinate conditions can be written in more
conventional form\footnote{The differential form (4.11c,d) of the
coordinate conditions may be integrated to yield (4.11a,b) either
by imposing appropriate boundary conditions or by a procedure
given in Appendix B of III.} by eliminating $\pi^T$ and $g_i$ via
(4.7):
$$%4.11c
\pi^{ii}\,\!_{,jj} - \pi^{ij}\,\!_{,ij} = 0
~~~~~~~~~~~~~~~~~~~~~~~~~~~~ \eqno{(4.11c)}
$$
$$%4.11d
g_{ij,j} = 0 \; . ~~~~~~~~~~~~~~ \eqno{(4.11d)}
$$
One can see that these coordinate conditions are acceptable by
looking at those of the field equations that involve $\pa_tg_i$
and $\pa_t\pi^T$ The linear part of equation (3.15a) gives, as the
equations for the longitudinal part of $g_{ij}$,
 \renewcommand{\theequation}{4.\arabic{equation}}
\setcounter{equation}{11}
 \be%4.12
 \pa_t (g_{i,j} + g_{j,i}) = N_{i,j} + N_{j,i}\; .
 \ee
 The Lagrange multipliers $N_i \equiv g_{0i}$ are functions determined
only when coordinate conditions are imposed and must vanish at
infinity where space is flat.  Inserting (4.11b) into (4.12)
gives, consistent with the boundary conditions, $N_i = 0$
everywhere. Similarly, from (3.15b), one has
 \be%4.13
 \pa_t [ - (1/2\nabla^2 ) \pi^T ] = N \; .
 \ee
Condition (4.11a) implies $N\equiv (-g_{00})^{-1/2}=1$, again
consistent with the required asymptotic limit.

Alternately, one can see that equations (4.11) are physically
appropriate coordinate conditions by a direct comparison with the
known results of linearized theory as discussed in I. Thus, as
mentioned above, ${\cal H}_{\rm lin} = - \nabla^2g^T$ and 
${\fT}^{0i}_{\rm lin} = - 2\nabla^2 (\pi^{iT} + \pi^L\,\!_{,i})$ are
the linearized theory's Hamiltonian and momentum densities and so
their coefficients in the generator (4.10) must be $\d t$ and $\d
x^i$ respectively, in order that the form (4.3), also obtained in
I, be reproduced.

{\bf 4.4. {\it Hamiltonian form and independent variables for
linearized theory.}}  Since the generator is now
 \be%4.14
G = \int d^3x \: [\pi^{ijTT} \,\d g_{ij}\,\!^{TT} - {\cal H}_{\rm
lin} (\pi^{ijTT}, \:  g_{ij}\,\!^{TT})\d t + {\fT}^{0i}_{\rm
lin} (\pi^{ijTT}, g_{ij}\,\!^{TT})\: \d x^i]
 \ee
the linearized theory has been put into canonical form, with
$g_{ij}\,\!^{TT}$ and $\pi^{ijTT}$ as the two canonically
conjugate pairs of variables.

We will now see the usefulness of the linearized theory in
suggesting the choice of canonical variables for the full theory.
Since the identification is made from the bilinear part of the
Lagrangian $\pi^{ij}\: \pa_t g_{ij}$, which is the same as for the
linearized theory, the greater complexity of the full theory, {\it
i.e.}, its self-interaction, is to be found only in the
non-linearity of the constraint equations. Even in the constraint
equations, the linearized theory will guide us in choosing $g^T$
and $\pi^i$ as the four extra momenta to be solved for.

{\bf 4.5. {\it Hamiltonian form and independent variables for the
full theory.}} The full theory can now easily be put into
canonical form. The generator of (4.10) is, of course, also
correct for the full theory since it comes from the bilinear part
of the Lagrangian. The constraint equations (3.15c) now read (in a
coordinate system to be specified shortly)
 $$ %4.15a
-\nabla^2 g^T = {\cal P}^0 [g_{ij}\,\!^{TT}, \: \pi^{ijTT}; \;
g^T, \: \pi^i ; \; g_i , \: \pi^T ]\eqno{(4.15a)}
$$
 $$ %4.15b
-2\nabla^2 (\pi^{iT} + \pi^L \,\!_{,i}) = {\cal P}^i
[g_{ij}\,\!^{TT}, \: \pi^{ijTT}; \; g^T, \: \pi^i ; \; g_i , \:
\pi^T ]~~~~~~~~~~~~~~ \eqno{(4.15b)}
$$
where ${\cal P}^\m$ are non-linear functions of $g_{ij}$ and
$\pi^{ij}$. One can again solve these (coupled) equations (at
least by a perturbation-iteration expansion) for $g^T$ and
$\pi^i$. Thus, one can again choose -- $\nabla^2g^T$ and
$-2\nabla^2 (\pi^{iT} + \pi^L\,\!_{,i})$ as the four extra momenta
to be eliminated. We denote the solutions of equations
 $$ %4.16a
-\nabla^2 g^T = -{\fT}^0\,\!_0 [g_{ij}\,\!^{TT}, \: \pi^{ijTT},
\;  g_i , \: \pi^T ]\eqno{(4.16a)}
$$
 $$ %4.16b
-2\nabla^2 (\pi^{iT} + \pi^L \,\!_{,i}) = -{\fT}^0\,\!_i
[g_{ij}\,\!^{TT}, \: \pi^{ijTT}, \;  g_i , \: \pi^T ] \; .
~~~~~~~~~~~~~ \eqno{(4.16b)}
$$
These equations are the counterpart of $p_{M+1} = -H$ in the
particle case.

As we have seen, the four constraint equations are maintained in
time as a consequence of the other field equations. Hence, after
inserting equations (4.16) into (3.15a,b), one finds that four of
these twelve (those for $\pa_t g^T$ and $\pa_t\pi^i$) are
``Bianchi" identities, leaving eight independent equations in the
twelve variables $g_{ij}\,\!^{TT}, \: \pi^{ijTT}, \;  g_i , \:
\pi^T$ and $N$, and $N_i$. These equations are linear in the time
derivatives of the first eight variables.

We now impose the coordinate conditions (4.11) which determine
$\pi^T$ and $g_i$.  The $\pa_t g_i$ and $\pa_t\pi^T$ equations
become determining equations for $N$ and $N_i$ [the full theory's
analog of (4.12) and (4.13)]. $N$ and $N_i$ are no longer 1 and 0
respectively, but now become specific functionals of
$g_{ij}\,\!^{TT}$ and $\pi^{ijTT}$, which could (in principle) be
calculated explicitly.\footnote{An example of such an explicit
determination of $N$ and $N_i$ for coordinate conditions (4.22) is
given in (7.7) and (7.9).}  In the last four equations, then, $N$
and $N_i$ may, in principle, be eliminated, leaving a system of
four equations involving only $g_{ij}\,\!^{TT}$ and $\pi^{ijTT}$,
and linear in their time derivatives.  We will now see that this
reduced system is in Hamiltonian form.

The generator (4.10) reduces to canonical form [with coordinate,
conditions (4.11) imposed and constraints (4.16) inserted]:
$$%4.17a
G = \int d^3x \, [ \pi^{ijTT} \; \d g_{ij}\,\!^{TT} + 
{\fT}^0\,\!_0 \: \d t + {\fT}^0\,\!_i \: \d x^i ] \eqno{(4.17a)}
 $$
while the Lagrangian now becomes
$$%4.17b
{\cal L} =  \pi^{ijTT} \; \pa_t g_{ij}\,\!^{TT} + {\fT}^0\,\!_0
\; .
 \eqno{(4.17b)}
 $$
It can be shown that the solutions ${\fT}^0\,\!_\m$ of the
constraint equations do not depend explicitly on the coordinates
$x^\m$ of (4.11) (see III). This is not unexpected, since the
variables $g_{ij}$ and $\pi^{ij}$ appearing on the right-hand side
of (4.15) do not depend explicitly on the coordinates in this
frame. (Thus, only $g_{i,j} = x^i\,\!_{,j} = \d^i\,\!_j$ and
$\pi^T = -2\nabla^2 t = 0$ appear in $g_{ij}$ and $\pi^{ij}$.)

{\bf 4.6. {\it Fundamental Poisson brackets and P.B.\ equations of
motion for general relativity.}} With the generator now in
canonical form, we can immediately write down the fundamental
equal time P.B.\ relations for $g_{ij}\,\!^{TT}$ and $\pi^{ijTT}$.
These are\footnote{The variables being used as the coordinates
$x^\m$ (4.11a,b) of course have vanishing P.B.\ with {\it all}
variables.}
$$%4.18a
[ g_{ij}\,\!^{TT} ({\bf x}), \: \pi^{mnTT}({\bf x}^\prime )] =
\d^{mn}\,\!_{ij} ({\bf x} -{\bf x}^\prime )~~~~~~~~~~~~~~~~~~
 \eqno{(4.18a)}
 $$
 $$%4.18b
[ g_{ij}\,\!^{TT} ({\bf x}), \: g_{mn}\,\!^{TT}({\bf x}^\prime )]
\; = 0   = [\pi^{ijTT} ({\bf x}), \: \pi^{mnTT} ({\bf x}^\prime
)]\; .
 \eqno{(4.18b)}
 $$
The $\d^{mn}\,\!_{ij} ({\bf x})$ in (4.18a) is a conventional
Dirac $\d$-function modified in such a way that the
transverse-traceless nature of the variables on the left-hand side
is not violated. Note that the definition of this modified
$\d$-function does not depend on the metric; it is symmetric,
transverse, and traceless on each pair of indices: \pagebreak
 $$%4.19a
 \d^{mn}\,\!_{ij} =  \d^{nm}\,\!_{ij} = \d^{mn}\,\!_{ji} =
 \d^{ij}\,\!_{mn}~~~~~
 \eqno{(4.19a)}
 $$
 $$%4.19b
 \d^{mm}\,\!_{ij} =  0 = \d^{mn}\,\!_{ii}
 \eqno{(4.19b)}
 $$
$$%4.19c
 \d^{mn}\,\!_{ij,j} =  0 \; .
 \eqno{(4.19c)}
 $$

From the form of equations (4.17), one also has the P.B.\
equations of motion:
$$%4.20a
 \pa_t g_{ij}\,\!^{TT} = [g_{ij}\,\!^{TT}, \: H] =
  \d H/\d\pi^{ijTT}~~~
 \eqno{(4.20a)}
 $$
$$%4.20b
 \pa_t \pi^{ijTT} = [\pi^{ijTT}, \: H] = -\d
 H/\d g_{ij}\,\!^{TT}
 \eqno{(4.20b)}
 $$
where $H \equiv - \int \: d^3x \:{\fT}^0\,\!_0$ is the
Hamiltonian. The last equalities in equations (4.20) follow from
equations (4.18). That (4.18) and (4.20) are consistent with the
Lagrangian equations obtained by varying (4.17b) is now
immediate.\footnote{The primary consistency criterion on the
canonical reduction is that equations (4.20) be identical to the
Einstein equations in the coordinate frame (4.11); this is
established in IIIA.} Corresponding to the time translation
equations (4.20), one also has for spatial displacements
$$%4.21a
 \pa_k g_{ij}\,\!^{TT} = [P_k , g_{ij}\,\!^{TT}] = -\d P_k/
\d\pi^{ijTT}
 \eqno{(4.21a)}
 $$
$$%4.21b
 \pa_k \pi^{ijTT} = [P_k , \pi^{ijTT}] = \d P_k/
\d g_{ij}\,\!^{TT}
 \eqno{(4.21b)}
 $$
where $P_k \equiv + \int {\fT}^0\,\!_k \: d^3x$ is the total
momentum operator. With the canonical momentum  $P_k\,\!^c \equiv
\int d^3x \: {\fT}^0\,\!_k\,\!^c \equiv -\int \: d^3x \:
\pi^{mnTT} g_{mn}\,\!^{TT}\,\!_{,k}$, equations (4.21) are
obvious. The quantities $P_k$ and $P_k\,\!^c$ actually coincide,
since their respective integrands differ by a divergence of
canonical variables, as shown in IIIA.

{\bf 4.7. {\it Alternate canonical form arising from different
coordinate conditions.}} The canonical form of equations (4.17)
is, of course, not unique. First one can make the usual type of
canonical transformations among the canonical variables
$g_{ij}\,\!^{TT}$  and $\pi^{ijTT}$, leaving the coordinate
conditions fixed. A more general class is available, however: the
generator (4.2) can also be reduced to canonical form with
coordinate conditions other than (4.11). That the reduction can
always be carried out is shown in Appendix A of III. An example
which will be of use later is given by the coordinate conditions
$$%4.22a
x^i = g_i - (1/4\nabla^2 ) g^T\,\!_{,i}~~~~~~~~~~
 \eqno{(4.22a)}
 $$
 $$%4.22b
t =  - (1/2\nabla^2 )  (\pi^T + \nabla^2 \pi^L )
 \eqno{(4.22b)}
 $$
 or, in differential form,
$$%4.22c
g_{ij,jkk} - \textstyle{\frac{1}{4}} \, g_{kj,kji} -
\textstyle{\frac{1}{4}}\, g_{jj,kki} = 0
 \eqno{(4.22c)}
 $$
$$%4.22d
~~~~~~~~~~~~~~~~~~~~~~~~~~~~~~~~~ \pi^{ii} = 0\; .
 \eqno{(4.22d)}
 $$
In this frame, the canonical variables are $g_{ij}\,\!^{TT}$  and
$\pi^{ijTT}$  (where the $^{TT}$ symbol now refers to the
coordinate system (4.22) and hence these canonical variables are
{\it different} from the set previously considered).  The 
${\fT}^0\,\!_\m$ in this frame are the solutions of the constraint
equations (4.16) for $\nabla^2 g^T$ and $-2\nabla^2 (\pi^{iT} +
\pi^L\,\!_{,i})$, which will be {\it different} functions of the
new canonical variables, since again $^T$ and $^L$ are referred to
(4.22) rather than (4.11). This frame is of interest since, for
$g_{ij}\,\!^{TT} = 0$ , the metric $g_{ij}$ reduces to the
isotropic form, {\it i.e.}, $g_{ij} = (1 + \frac{1}{2} \,
g^T)\d_{ij}$. We will discuss, in Section 5, a preferred class of
physically equivalent frames which includes the two given here.

 \renewcommand{\theequation}{4.\arabic{equation}}
\setcounter{equation}{22}

It is perhaps worth noting that, with the coordinates of (4.22),
the orthogonal decomposition of $g_{ij}$ takes on the simple form
 \be%4.23
g_{ij} = g_{ij}\,\!^{TT} + \d_{ij} (1+ \textstyle{\frac{1}{2}}\:
g^T \, [g_{mn}\,\!^{TT}, \: \pi^{mnTT}]) \; .
 \ee
Thus, the dynamical aspects of the theory are, as expected, to be
found in the deviation of the metric from its flat space value. By
tracing (4.23), one obtains a local relation giving the canonical
variable $g_{ij}\,\!^{TT}$ at a point in this coordinate system in
terms of $g_{ij}$  at that point: $g_{ij}\,\!^{TT} = g_{ij} -
\frac{1}{3} \, \d_{ij} g_{kk}$. The $g^T$ term, as we have seen,
is directly given by the Hamiltonian density, and, as will be
shown in the next section, carries the asymptotic (Newtonian)
$m/r$ term in the metric.

\noindent{\bf 7-5. Energy and Radiation}

{\bf 5.1. {\it Expressions for the energy and momentum $P_\m$ of
the gravitational field.}} The canonical formalism developed in
the previous section has brought out the formal features of
general relativity which have their counterpart in usual Lorentz
covariant field theories. As a consequence, the physical
interpretation of the gravitational field may be carried out in
terms of the same quantities that characterize other fields, {\it
e.g.}, energy, momentum, radiation flux (Poynting vector).

\renewcommand{\theequation}{5.\arabic{equation}}
\setcounter{equation}{0}

The energy $E$ of the gravitational field is just the numerical
value of the Hamiltonian for a particular solution of the field
equations. In obtaining this numerical value, the form of the
Hamiltonian as a function of the canonical variables is
irrelevant, and one may make use of the equation, 
${\fT}^0\,\!_0 = \nabla^2 g^T$ to express $E$ as a surface integral.
One has then\footnote{It should be emphasized that, while the {\it
energy} and {\it momentum} are indeed divergences, the integrands
in the {\it Hamiltonian} and in the space translation generators
${\fT}^0\,\!_\m [g^{TT}, \, \pi^{TT}]$ are {\it not}
divergences when expressed as functions of the canonical
variables.}
\begin{eqnarray}%5.1
P^0 \equiv E = -\int \: d^3{\bf r} \: \nabla^2 g^T = & =& - \oint
\, dS_i \, g^T\,\!_{,i} \nonumber \\
& = & \oint \, dS_i \, (g_{ij,,j} - g_{jj,i})
 \eea
where $dS_i$ is the two-dimensional surface element at spatial
infinity. Similarly, the total field momentum $P_i$ may be written
as
\be%5.2
P^i = - 2 \, \oint \, (\pi^i\,\!_{,j} + \pi^j\,\!_{,i})dS_j = -2
\, \oint \, \pi^{ij} \, dS_j \; .
 \ee
In (5.1) and (5.2) we have assumed, as throughout, that the
coordinate system is asymptotically rectangular.  The basic
requirement for an energy to be at all defined is, of course, that
space-time become flat at spatial infinity (so that the integrals
in (5.1,2) are well defined).  The rectangularity requirement may
then be conveniently imposed to avoid unnecessarily complicated
expressions.

{\bf 5.2. {\it Coordinate invariance of energy-momentum
expression; Lorentz four-vector nature of $P_\m$.}} The usefulness
of formulas (5.1) and (5.2) is enhanced by the fact that, in
evaluating the surface integrals at infinity, it is not necessary
to employ there the original canonical coordinate frame. At
spatial infinity, where the metric approaches the Lorentz value,
coordinate transformations, which preserve this boundary
condition, must approach the identity transformation (excluding,
for the moment, Lorentz transformations at infinity).  For the
transformation $\bar{x}^\m = x^\m - \xi^\m$ then, one has
$\xi^\m\,\!_{,\n} \rightarrow 0$ (since only $\pa\bar{x}^\m /\pa
x^\n$ occurs in the transformation laws of $g_\mn$ and
$\G_\rho\,\!^\a\,\!_\s )$. Let us first keep only the linear terms
in the transformation:
$$%5.3a
\bar{g}_{ij} = g_{ij} + \xi^i\,\!_{,j} + \xi^j\,\!_{,i}~~~~~~~~~
 \eqno{(5.3a)}
 $$
$$%5.3b
\bar{\pi}^{ij} = \pi^{ij} - (\d_{ij}\xi^0\,\!_{,ll} -
\xi^0\,\!_{,ij})\; .
 \eqno{(5.3b)}
 $$
For the entire formalism, it is necessary that $g_\mn - \eta_\mn$
and $\G_\rho\,\!^\a\,\!_\s$ fall off at least as $\sim 1/r$ at
spatial infinity.  Hence, $\xi^\m\,\!_{,\n}$ , must have a similar
behavior.  As can be seen from (4.5), these transformations affect
only $g_i$ and $\pi^T$ leaving $g^T\,\!_{ij}, \:, \pi^i\,\!_{,j}$,
and the canonical variables invariant to $O(1/r)$.  However, in
(5.1) and (5.2), one needs to know $g^T\,\!_{,i}$  and $
\pi^i\,\!_{,j}$ , to $O(1/r^2)$, so that one must also investigate
the quadratic terms in the transformation law.  These present a
problem only if a derivative acting  on $\xi^\m\,\!_{,\n}$ does
not alter its order, {\it i.e.}, if one has a ``coordinate wave"
$\xi^\m\,\!_{,\n} \sim e^{ipx}/r$ whose derivatives are again
$\sim 1/r$. A detailed investigation (IVC) shows that expressions
(5.1) and (5.2) for $P^\m$ {\it are} invariant under {\it all}
transformations preserving $\bar{g}_\mn - \eta_\mn \sim 1/r$,
provided one averages the surface integrals over the oscillatory
terms. The prescription of averaging is justified, since the
energy and momentum can be defined directly from the leading $\sim
1/r$ terms of $g^T$ and $\pi^i$, in the coordinate frame (4.11).
[For instance, from (4.16a), one sees that $g^T =
(1/\nabla^2){\fT}^0\,\!_0 \sim E/4\pi r$.] However, as shown in
IVC, these $1/r$ terms are in fact coordinate invariant and the
averaged $P^\m$ integrals agree with them.  Thus, by either
prescription, our formula for $P_\m$ is unaffected by changes of
coordinates not involving Lorentz transformations at spatial
infinity.

The definition of energy and momentum, as given by the canonical
formalism, may be shown to coincide with that arising from the
stress energy pseudotensor of Landau and Lifshitz.  In this
definition, the energy and momentum can also be given in surface
integral form. (Here, too, one must average over any oscillatory
terms to obtain a coordinate-independent result.) The linear terms
in this pseudotensor are then just the surface integrals in (5.1)
and (5.2), while the higher powers are negligible at infinity
(using the averaging).  This agreement also enables us to note
directly that $P^\m$ behaves as a Lorentz vector under rigid
Lorentz transformations of the coordinate system, since this
property is manifest for the pseudotensor expression. That the
energy of a system should be defined from the asymptotic metric
alone follows from the equivalence principle, the total
gravitational mass being measurable by the acceleration of a test
particle at infinity.

Since $P^\m$ is constant, due to the fact that ${\fT}^0\,\!_\m$
does not depend on $x^\m$, explicitly, it can be evaluated at any
given time. For this purpose, one should need only those initial
Cauchy data required to specify the state of the system uniquely.
Prior to the imposition of coordinate conditions, these are
$g_{ij}$ and $\pi^{ij}$ (and not, for example, $g_{0\mu}$) as we
have seen in Section 3. The last members of (5.1) and (5.2) are so
expressed. When any (asymptotically rectangular) coordinate
conditions are imposed, one needs only the two pairs of canonical
variables of that frame to specify the state of the system and, of
course, to calculate $P^\m$ (using the formula $P_\m = \int \,
d^3r \, {\fT}^0\,\!_\m$).

{\bf 5.3. {\it Conditions for equality of Hamiltonian and
energy---``Heisenberg representation."}} As was discussed in the
previous section, there are an infinite number of coordinate
conditions that may be invoked to put the theory in canonical
form.  The question therefore arises as to whether the above
physical results are affected by the coordinate conditions chosen.
While, for example, we have seen that the $P_\m$ of (5.1) and
(5.2) is invariant under change of frame, this does not prove that
different canonical forms, arising from different frames, will
yield the same integral for $P_\m$.  Indeed, not every canonical
form of the theory has the same value of the Hamiltonian, as one
is free to perform Hamilton--Jacobi (H.-J.), or partial H.-J.,
transformations.  One must therefore. take as primary, frames and
canonical variables which are in the ``Heisenberg representation."
More precisely, we require that in the expression giving the full
metric $g_\mn$ at any time in terms of the canonical variables at
that time, there be no explicit coordinate dependence.  The
preferred role is given to the metric in this requirement since
basic measurements (rods and clocks) refer to $g_\mn$ directly.
Similarly,in standard particle mechanics, the directly measured
quantities (which appear in the original Lagrangian) are the basic
``Heisenberg" variables, and one requires that any other
``Heisenberg" set be related to them in a time-independent
fashion. More generally, even if the original Lagrangian (which
always defines the Heisenberg representation) is not in canonical
form, the same requirement is to be imposed.  For relativity, the
basic variables in the Einstein Lagrangian are  $g_\mn$. (Note
that the question of whether a given canonical form is Heisenberg
thus does not require a comparison with some particular canonical
form, and also that each of the two forms in Section 4 is indeed
Heisenberg.) With this criterion, it is clear that the Hamiltonian
densities will also have no explicit coordinate dependence, since
the canonical equations can be obtained by substituting $g_\mn
[g^{\rm can}, \: \pi^{\rm can}]$ into the original field
equations.  This step produces no explicit coordinate dependence
in the Heisenberg representation. It can be shown (see IVA) that
all Heisenberg forms are linked by coordinate transformations of
the type $\bar{x}^\m = x^\m + f^\m [g^c (x), \; \pi^c(x)]$ where
the $f^\m$ do not depend on $x^\m$ explicitly. Thus the relation
between Heisenberg frames is determined by the intrinsic geometry
and does not contain functions of the coordinates, which are
independent of the physics of the state. As a result of this form
of the transformations, one finds that $P_\m$ is numerically the
same in any Heisenberg frame for a given physical situation. Thus
(5.1) gives correctly the numerical value of the Hamiltonian of
any Heisenberg frame and could have been derived in any Heisenberg
canonical form.

{\bf 5.4. {\it ``Waves" as excitations of canonical variables.}}
Excitations of the canonical modes of a field provide a definition
of what one calls waves.  This is, of course, the same definition
as that given in electrodynamics.  In electrodynamics, the gauge
invariance partly obscures the fact that only the transverse parts
${\bf A}^T$ and ${\bf E}^T$ of the vector potential and
electric field need be examined to recognize waves.  These
variables, as we saw in Section 4, are just the canonical ones for
the Maxwell field. Correspondingly, in general relativity we may
utilize the non-vanishing of the canonical variables as the
criterion of the existence of waves. As in electrodynamics, when
sources (among which are included the non-linear
self-interactions) are present, the radiation and induction
effects can be meaningfully separated only in the ``wave zone";
but also as in electrodynamics, the wave concept can be employed
consistently nearer the sources where it is of heuristic value,
although the definition is somewhat arbitrary.

\renewcommand{\theequation}{5.\arabic{equation}}
\setcounter{equation}{3}

An example of an application of the wave criteria is furnished by
the time-symmetric case ($\pi^{ij} = 0$ initially).  In the frame
of (4.22), it was seen that $\pi_{ij}\,\!^{TT} = 0$  if, and only
if, the metric is isotropic, $g_{ij} = (1 + \frac{1}{2} \,
g^T)\d_{ij}$. One need not explicitly transform to this frame to
verify whether the canonical variables vanish in it; it is
sufficient to examine the three-tensor (see, for example,
Eisenhart, 1949, Section 28):
\be%5.4
R_{ijk} \equiv R_{ij|k} - R_{ik|j} + \textstyle{\frac{1}{4}} \,
(g_{ik} R_{,j} - g_{ij} R_{,k})
 \ee
which vanishes if, and only if, $g_{ij}\,\!^{TT}$ vanishes in the
frame\footnote{In terms of the canonical formalism, the
time-symmetric situation $\pi^{ij}=0$ can be viewed as the
requirement that excitations in the canonical momentum variable
$\pi^{ijTT}$ vanish. This follows from the fact that if
$\pi^{ijTT} = 0$, then $\pi^i = 0$ by the constraint equations
(3.15c); consequently, $\pi^T = 0$ in the frame of (4.22). Thus,
in this frame, $\pi^{ijTT} = 0$ implies $\pi^{ij} = 0$.} of
(4.22). A class of time-symmetric waves is furnished by the
initial conditions recently discussed by Brill (1959).

{\bf 5.5. {\it Radiation and definition of wave-zone.}}  The
concept of radiation embodies the idea of field excitations
escaping to infinity and propagating independently of their
sources.  In linear field theories, the region in which this
occurs (the wave zone) is simply one which is many wavelengths
away from the source.  In general relativity, for strong fields,
nonlinear effects are important, and act as effective sources, so
that merely going beyond the matter sources is not, in general,
adequate to define free radiation.  To have a wave zone situation,
one must set up further conditions such that the non-linear
self-interaction terms do not influence the propagation of the
waves.  Such further conditions are also necessary in quantum
electrodynamics due to vacuum polarization. The latter brings in
effective non-linear contributions to the Maxwell action, so that
in the presence of an arbitrary external electromagnetic field,
the self-coupling can produce distortion of the ``waves"
(Delbr\"{u}ck scattering); again, scattering of two photons can
occur, even in the absence of external fields, thus affecting
superposition, that is, free propagation.  The usual definitions
of radiation would thus fail in such situations and, in fact, the
``wave zone" is then defined only when the self-coupling is
negligible.  One must therefore require that in the wave zone, the
various field amplitudes be small when the theory is non-linear.
On the other hand, the process of going to the asymptotic wave
zone region is not an {\it a priori} equivalent to taking the
linearized approximation. In general relativity, we have
previously seen that $(g^T,\:\pi^i) \sim P^\m /r$ asymptotically,
where $P^\m$ is clearly determined by the interior region.  Thus,
components of the metric other than the wave modes survive
asymptotically, and in fact may be comparable in size to the wave
modes.

To define the wave zone, we consider a general situation in which
the gravitational canonical modes behave as $fe^{ikx}/r$
asymptotically in some region (with wave numbers $k$ up to some
maximum $k_{\rm max}$); beyond some point (the wave front) they
are assumed to vanish very rapidly. (This last assumption is
needed to make the energy of the waves finite.) The first wave
zone requirement is the familiar one that $kr >\!\!\!> 1$; this
implies that gradients and time derivatives acting on the
canonical modes are also $\sim 1/r$. Second, we demand that for
all components of the metric, $|g_\mn - \eta_\mn | \sim a/r
<\!\!\!< 1$. This criterion can always be met (for radiation
escaping to infinity) by taking $r$ large enough, and waiting for
the wave to reach this distance. This condition implies that, not
only is the wave disturbance weak $(f/r <\!\!\!< 1)$, but also the
Newtonian-like parts of the field ({\it e.g.}, $g^T$) are small
$(P^\m/r <\!\!\!< 1)$. It insures that quadratic (and higher)
terms in $g_\mn - \eta_\mn$, are negligible compared to the linear
ones. Finally, we impose the condition that $|\pa g/\pa
(kx)|^2<\!\!\!< |g - \eta |$. In terms of $k_{\rm min}$, the
minimum frequency (or wave number) to be considered (with, of
course, $k_{\rm min} r
>\!\!\!> 1$), one has $k_{\rm min} r >\!\!\!> k_{\rm max} (a/r)^{1/2}$.
Again, for large enough $r$, this inequality can always be
achieved. [For fixed $r$, it sets a lower bound on frequencies
which may be treated in the wave zone.] The purpose of this
requirement is to make the non-linear structures containing
gradients also small compared to the leading linear terms.  The
last two conditions together are arranged to guarantee the absence
of all self-interactions in the wave zone.

{\bf 5.6. {\it General structure of gravitational radiation;
superposition, coordinate-independence.}} With the above
definition of the wave zone, one finds (see IVB) that the field
equations (3.15a,b) reduce to
$$%5.5a
\pa_t g_{ij} = 2 (\pi^{ij} - \textstyle{\frac{1}{2}} \, \d_{ij}
\pi^{ll}) + N_{i,j} + N_{j,i} + O(1/r^2 )
 \eqno{(5.5a)}
 $$
$$%5.5b
\pa_t \pi^{ij} = \textstyle{\frac{1}{2}} (g_{ij,kk} + g_{kk,ij} -
g_{ik,kj} - g_{jk,ki})~~~~~~~~~~
$$
$$
~~~~~~~~~~ - \textstyle{\frac{1}{2}} \d_{ij}(g_{kk,ll} -
g_{kl,kl}) - (\d_{ij} N_{,kk} - N_{,ij} )
$$
$$
+ \; O(1/r^2)~~~~~~~~~~~~~~~~~~~~~~~~~~~
 \eqno{(5.5b)}
 $$
where  $O(1/r^2)$ is much less than the leading terms.  Using a
result of IVB that the orthogonal parts of $\sim 1/r$ and  $\sim
1/r^2$ structures are also at most  $\sim 1/r$  and  $\sim 1/r^2$
respectively, one finds that
$$
\pa_t g_{ij}\,\!^{TT} = 2\pi^{ijTT}\eqno{(5.6a)}
$$
$$
\pa_t \pi^{ijTT} = \textstyle{\frac{1}{2}} \nabla^2
g_{ij}\,\!^{TT} \eqno{(5.6b)}
$$
to the order $1/r$. Thus the {\it rigorous} dynamical modes obey
the linearized theory's equations in the wave zone. Equations
(5.6) are, of course, source-free, so that the radiation
propagates independently of its origin, and without any
self-interaction or dependence on the Newtonian-like parts $g^T$
and $\pi^i$ of the field.  The flat-space wave equation with
constant coefficients, $\Box^2\,\!_{\rm flat}g_{ij}\,\!^{TT} = 0$,
represented by (5.6), indicates the absence of curvature effects
on the radiation. Further, there are no coordinate effects left,
since the coordinate-dependent components $g_i,\;\pi^T$, and
$g_{0\m}$, have disappeared.

We have seen previously that $g_{ij}\,\!^{TT}$ and $\pi^{ijTT}$
are invariant to the order $1/r$ under coordinate transformations
leaving the metric asymptotically flat and not involving Lorentz
transformations. Thus, in a fixed Lorentz frame, $g^{TT}$ and
$\pi^{TT}$ represent a coordinate-independent description of the
radiation. The detailed analysis of IVB shows how any coordinate
waves which may be present can be identified unambiguously.

The above derivation may be carried through equally well in the
presence of coupling to matter ({\it e.g.}, the Maxwell field),
provided of course, the matter system is bounded within the
interior region (and has finite energy content).  Under these
circumstances, any gravitational waves in the wave zone are still
independent of the matter sources. If, however, electromagnetic
radiation is also propagating in the wave zone with small enough
amplitude (and $\sim 1/r$), the gravitational radiation is still
free, being unaffected to $\sim 1/r$ by the electromagnetic waves,
since the latter couple quadratically   ($\sim 1/r^2$)  in the
field equations. Hence, in the common wave zone, the systems are
correctly independent.

\renewcommand{\theequation}{5.\arabic{equation}}
\setcounter{equation}{6}

{\bf 5.7. {\it Measurables in the wave zone: the Poynting vector,
radiation amplitudes.  Relation to curvature tensor.}} Since the
dynamics of the rigorous canonical modes is identical, in the wave
zone, to that given by linearized theory, any radiation arising
from strong interior fields can be precisely simulated by a
corresponding solution of the linearized theory with sources. Such
an analog metric can be made everywhere weak, so that linearized
theory is everywhere valid, and one may arrange the sources so
that in the wave zone, the correct $g^T$ and $\pi^i$ result. An
absorber of gravitational radiation in the wave zone clearly
cannot distinguish between the true and analog situations.  One is
therefore free to use the Poynting vector of linearized theory to
measure the energy flux.  From the symmetric stress-tensor of 
I, one finds straightforwardly that
\be%5.7
T^{i0}_{{\rm lin}} = \pi^{lmTT} (2g^{TT}_{li,m} - g^{TT}_{lm,i})
\ee
 in the coordinate frame of (4.11). However, it is easy to
see that the general form of  $T^{i0}_{{\rm lin}}$, is
coordinate-invariant through terms of $0(1/r^2)$, as is the
right-hand side of (5.7). Hence (5.7) provides an invariant
formula for the wave-zone Poynting vector.  We remark that, as in
electrodynamics, the physical part of the Poynting vector is
obtained by averaging over oscillatory terms.  This averaging is
of use in proving that $T^{i0}_{\rm lin}$ is
coordinate-independent, since it also eliminates ``coordinate
wave" effects.  It may also be noted that, in (5.7), no terms
involving $g^T, \; \pi^i$ have survived to $O(1/r^2)$, showing
that the Newtonian-like parts do not affect the energy flux of
radiation. Similarly, in electrodynamics ${\bf E}^L$ does not
contribute to ${\bf E \mbox{\boldmath{$\times$}} B}$ since  ${\bf
E}^L \sim 1/r^2$.

The canonical variables $g_{ij}\,\!^{TT}$ and $\pi^{ijTT}$ are, in
general, non-local functions of the metric and its first
derivatives. However, it is possible to obtain their $1/r$ terms
by measurements entirely within the wave zone.  From the condition
$kr >\!\!\!> 1$, it is possible to isolate sufficiently accurately
the $k$ Fourier component of $g_{ij}$ and $\pa_t g_{ij}$, and to
obtain $g_{ij}\,\!^{TT}(k), \;\pi^{ijTT}(k)$  algebraically from
these. Equivalently, one may measure the $k$-component of the
curvature tensor $R_\a\,\!^\m\,\!_{\n\b}$ to obtain these physical
quantities since it may be shown that in the wave zone
$$%5.8
g_{ij}\,\!^{TT} = (2/{\bf k}^2) R_i\,\!^m\,\!_{mj}~\eqno{(5.8a)}
$$
$$
\pi_{ijTT} = -(ik_\ell{\bf k}^2) R_i\,\!^0\,\!_{lj} \; .
 \eqno{(5.8b)}
$$
Such measurements find a parallel in Maxwell theory, where ${\bf
E}^T, \;{\bf A}^T $ are found from {\bf E} and {\bf B} by
$$%5.9
{\bf E}^T = {\bf E} - {\bf k}[({\bf k \cdot E})/{\bf k}^2]
\eqno{(5.9a)}
$$
$$
{\bf A}^T = i({\bf k} \mbox{\boldmath$\times$} {\bf B})/{\bf k}^2
\; .~~~~~~~
 \eqno{(5.9b)}
$$
In fact, electromagnetic wave measurements are commonly of this
type.

It should be noted that just as the definition of electromagnetic
radiation is not restricted to outgoing radiation (but may include
mixtures of outgoing and incoming waves), this also is true of the
above definition of gravitational radiation.  For the purely
outgoing case, the $1/r$ part of the curvature tensor is of type
II---Null in Petrov classification (Petrov, 1954, and Pirani,
1957). We have seen in this section that the leading $(\sim 1/r)$
physical terms in the field past the wave front are $g^T$ and
$\pi^i$, and that these yield invariantly the energy-momentum
vector of the system. In the wave zone itself, the dynamical modes
specifying the radiation, as well as the Poynting vector, could be
obtained invariantly.  Actually, a good deal of detailed
information about the strong fields in interior regions is
obtainable by purely asymptotic (``$S$-matrix")
measurements,\footnote{J. Plebanski (private communication) has
shown how properties of orbits of test particles may be examined
by purely asymptotic measurements (using geodesic projections)
even if the orbits remain entirely within the strong field
interior region.} so that a large portion, at least, of the
description of a system is manifestly frame-independent, and can
be gotten without the need for introducing apparatus into strong
field regions and hence having to analyze its behavior there.

\noindent{\bf 6. Extension to Coupling with Other Systems}

\renewcommand{\theequation}{6.\arabic{equation}}
\setcounter{equation}{0}

{\bf 6.1. {\it Lagrangian for coupled Einstein-Maxwell-charged
particle system.}} Up to now we have considered the free
gravitational field, sketching in the process the main physical
features which have emerged there.  The extension to coupling
enables us to verify that these characteristics remain unchanged
and provides a basis for the self-energy calculations of the next
section.  The analysis is completely analogous to the discussion
of the free field and so we limit ourselves to a brief outline for
the case of coupling to electrodynamics with point charges (more
details may be found in V and VA). The addition to the Einstein
Lagrangian is
 \bea%6.1
 \lefteqn{{\cal L}_M =  \textstyle{\frac{1}{2}} \: (A_{\m ,\n} - A_{\n
 ,\m}) {\fF}^\mn + \textstyle{\frac{1}{4}}\: (-^4g)^{1/2} \:
{\fF}^\mn{\fF}^{\a\b} \;  ^4 {\fg}_{\m\a}\, ^4{\fg}_{\n\b}}\nonumber \\
&& + \int ds \: e(dx^\m (s)/ds)A_\m (x)\: \d^4 (x - x(s))\nonumber \\
&& + \int ds \: \{\pi_\m (dx^\m /ds) - \textstyle{\frac{1}{2}}\,
\l^\prime (s) (\pi_\m\pi_\n\;^4  g^\mn + m_0\,\!^2)\}
\d^4(x-x(s))\;.
 \eea
 We are here employing a first-order formalism for the particle as
well as for the Maxwell field.  The Maxwell part of ${\cal L}_M$
is the straightforward covariant generalization of (3.5), with the
field strength ${\fF}^\mn$ now a tensor density.  Thus, $A_\m$
and ${\fF}^\mn$ are varied separately, as are $x^\m (s)$, the
mechanical momentum $\pi_\m (s)$, and the Lagrange multiplier
$\l^\prime (s)$. The particle Lagrangian is in parameterized form
(with arbitrary parameter\footnote{Equation (6.1) is covariant
against any reparameterization $\bar s = \bar s(s)$ with
$\l^\prime$ transforming as a ``vector": $\bar \l ^\prime =  \l
^\prime (ds/d\bar s)$ (just as $N$ does in Section 2).} $s$), as
is required for manifest covariance.  The $\d^4$-function is a
scalar density, invariantly defined in a metric-independent way by
$\int \d^4 (x) d^4x = 1$; the $\d^3$-function is similarly defined
in three-space, {\it i.e.}, $\int \d^3 ({\bf r}) d^3r = 1$.
Introducing the notation ${\cal E}^i \equiv {\fF}^{0i}$ and
${\cal B}^i \equiv \textstyle{\frac{1}{2}} \: \e^{ijk} (A_{k,j}
-A_{j,k})$ (where $\e^{ijk} = \e_{ijk} = 0, \; \pm 1$ is totally
antisymmetric) and using the gravitational variables of Section 2,
one obtains 
%\newpage
 \bea%6.2
 \lefteqn{{\cal L}_M = A_i \: \d_0{\cal E}^i + \d^3 ({\bf r} - {\bf
 r}(t))[(\pi_i (t) + eA_i)(dx^i /dt) + \pi^0]}\nonumber \\
 && + \; A_0 [e \:\d^3 ({\bf r} - {\bf r}(t)) - {\cal
 E}^i\,\!_{,i}]\nonumber \\
 && - \textstyle{\frac{1}{2}}\, \l [g^{ij}\pi_i\pi_j - N^{-2} (\pi_0
 - N^i\pi_i)^2 + m_0\,\!^2] \d^3 ({\bf r} - {\bf r}(t))\nonumber \\
 && - \textstyle{\frac{1}{2}} \, Ng^{-1/2} g_{ij} [ {\cal E}^i{\cal
 E}^j +{\cal B}^i{\cal B}^j] + N^i [\e_{ijk} {\cal E}^j{\cal
 B}^k]\; .
 \eea
We have also performed the $s$ integration which led to the
replacement of $\l^\prime$ by\\ 
$\l \equiv [\l^\prime
(s)(ds/dx^0(s)]_{x{^0}(s=t)}$. It is convenient next to the
eliminate the non-gravitational constraints obtained from varying
$A_0$ and $\l$. These are:
$$%6.3a
{\cal E}^i\,\!_{,i} = e \: \d^3({\bf r} - {\bf r}(t))
\eqno{(6.3a)}
$$
$$%6.4a
\pi^\m\pi_\m + m_0\,\!^2 = 0 \eqno{(6.4a)}
$$
whose solutions are
$$%6.3b
{\cal E}^{iL} = -\nabla (e/4\pi |{\bf r} - {\bf r}(t)|)~~~~~~~
~~~~~~~~ \eqno{(6.3b)}
$$
$$%6.4b
\pi_0 = N^i\pi_i - N (g^{ij}\pi_i\pi_j +  m_0\,\!^2)^{1/2}\; .
\eqno{(6.4b)}
$$
Here, we have again used the orthogonal breakup of ${\cal E}^i$ in
the same flat-space sense as in Section 4.  Therefore, ${\cal
E}^i$ is to be regarded subsequently as an abbreviation for ${\cal
E}^{iT} +{\cal E}^{iL}$, with ${\cal E}^{iL}$  expressed from
(6.3b).

\renewcommand{\theequation}{6.\arabic{equation}}
\setcounter{equation}{4}

{\bf 6.2. {\it Canonical reduction of coupled system.}} At this
stage,the matter Langrangian is in the reduced form
 \bea
 \lefteqn{{\cal L}_M = (-{\cal E}^{iT}) \, \pa_0 A_i\,\!^T + [p_i
 \, dx^i/dt] \d^3 ({\bf r} - {\bf r}(t))}\nonumber \\
&& - \;\textstyle{\frac{1}{2}}\, Ng^{-1/2} g_{ij} [ {\cal E}^i
{\cal E}^j +{\cal B}^i {\cal B}^j]\nonumber \\
&& - \; N[g^{ij} (p_i - eA_i\,\!^T)(p_j - eA_j\,\!^T) +
m_0\,\!^2]^{1/2} \: \d^3 ({\bf r} - {\bf r}(t))\nonumber \\
&& + \; N^i [\e_{ijk} {\cal E}^j{\cal B}^k + (p_i - eA_i\,\!^T)
\d^3 ({\bf r} - {\bf r}(t))]
 \eea
where $p_i \equiv \pi_i + eA_i\,\!^T$. The first two terms of
${\cal L}_M$ are in the standard $\S p\dot{q}$ form and thus
($A_i\,\!^T , - {\cal E}^{iT}$) are the canonical variables for
the electro-magnetic field, while $p_i(t),\; x^i(t)$ are those of
the particle.  Note that the variables $A_i, \; {\cal E}^i,\;p_i$
or $x^i$, which arose naturally in obtaining canonical form, are
also the appropriate variables from the geometrical (Cauchy
problem) considerations of Section 3. Thus, $A_i, \; p_i, \;
\pi_i$ are covariant spatial components of the corresponding
four-vectors, while $x^i(t)$ locates the particle within the
three-surface.  The three-vector density character of ${\cal E}^i$
is established from its relation to the covariant spatial
components of the four-dimensional dual $^*\!\!F_\mn$ of the field
strength:
$$
{\cal E}^i = \textstyle{\frac{1}{2}} \: \e^{ijk} \: ^*\!\!F_{jk}
\equiv \textstyle{\frac{1}{2}} \: \e^{ijk} (
\textstyle{\frac{1}{2}} \: \e_{jk\mn} {\fF}^\mn )
$$
The total Lagrangian, which is the sum of  ${\cal L}_M$ and the
gravitational part (3.13), now has precisely the parameterized
form of (2.3), as a consequence of the general covariance of each
part. The coefficients of $N$ and $N^i$ now yield the extended
constraint equations
$$%6.6a
gR +\textstyle{\frac{1}{2}} \, \pi^2 - \pi^{ij}\pi_{ij} = \sqrt{g}
\: \d^3 ({\bf r} - {\bf r}(t))[g^{ij} (p_i - eA_i\,\!^T)(p_j -
eA_j\,\!^T) + m_0\,\!^2)]^{1/2} +\textstyle{\frac{1}{2}}\: g_{ij}
[ {\cal E}^i{\cal E}^j + {\cal B}^i{\cal B}^j]
 \eqno{(6.6a)}
$$
$$%6.6b
- 2\pi_{ij}\,\!^{|j} = \d^3 ({\bf r} - {\bf r}(t))(p_i -
eA_i\,\!^T) + \e_{ijk} {\cal E}^j{\cal B}^k \; . \eqno{(6.6b)}
$$
Equations (6.6) can again be solved for
$$
\nabla^2g^T = {\fT}^0\,\!_0
$$
 and
$$
-2\nabla^2 (\pi^{iT} + \pi^i\,\!_{,i}) = {\fT}^0\,\!_i \; .
$$
When the coordinate conditions (4.11) are again imposed, the
theory is in canonical form with the {\it same} gravitational
canonical variables.  One sees most clearly now how the
gravitational canonical variables characterize independent
gravitational field excitations.  All the other components of the
metric, {\it e.g.}, $g^T$ and $N,\; N^i$ depend on the
non-gravitational variables as well. In contrast, only
$g_{ij}\,\!^{TT}$ and $\pi^{ijTT}$ can be specified initially,
irrespective of the excitations of the matter system.

\renewcommand{\theequation}{6.\arabic{equation}}
\setcounter{equation}{6}

The quantity $-{\fT}^0\,\!_0$ now represents the Hamiltonian
density of the total system\footnote{ln the flat space limit, this
Hamiltonian density is correctly the usual one for
electrodynamics.} and depends only on the canonical variables of
all three parts. Further, the expressions (5.1) and (5.2) now give
the energy-momentum of the total system. The fact that the total
energy of the interacting system is obtainable from purely metric
quantities is analogous to the expression in electrodynamics of
the total charge in terms of the integral of the longitudinal
electric field.  It is worth noting that the choice of coordinate
conditions is unaffected by coupling. Finally, we record the
equations of motion of the gravitational field including coupling.
The extension of the constraint equations (3.15c) has been given
by (6.6) above; the $\pa_tg_{ij}$ equations (3.15a) are unchanged,
since they are of the type $m\dot x =p$ (namely, they are defining
equations for Christoffel symbols). However, one must now add a
term $\frac{1}{2} \, N{\fT}_M{^{ij}}$ to the right-hand side of
(3.15b) for $\pa_t\pi^{ij}$, where ${\fT}_M{^{ij}}$ is the
symmetric spatial stress-energy tensor density of the matter:
 \bea%6.7
 {\fT}_M{^{ij}} & \equiv & g^{ij}g^{jk} \; \sqrt{g} \;
 ^4{\fT}_{Mlk} \nonumber \\
& = & \d^3 ({\bf r} - {\bf r}(t))(p^i - eA^{iT})(p^j -
eA^{jT})[({\bf p} - e{\bf A}^T)^2 + m_0\,\!^2]^{-1/2} \nonumber \\
&& \mbox{\,}+ g^{-1/2}[\textstyle{\frac{1}{2}}\: g^{ij} ({\cal E}^m{\cal E}_m
+{\cal B}^m{\cal B}_m) - {\cal E}^i{\cal E}^j - {\cal B}^i{\cal
B}^j]\; .
 \eea

\renewcommand{\theequation}{7.\arabic{equation}}
\setcounter{equation}{0}

\noindent{\bf 7-7. Static Self-Energies}

{\bf 7.1. {\it Physical basis for finiteness of self-energy when
gravitation is included.}} The canonical formalism developed for
coupled systems in the previous section allows one to define pure
particle states by the vanishing of the canonical variables
referring to field excitations.  For particles at rest, therefore,
the energies of such states are just the total rest and
interaction energies, which include, of course, their
self-energies.

The aim of classical point electron theory, since its inception,
has been to obtain a finite, model-independent electromagnetic
self-energy, and if possible, to dispense with mechanical mass
altogether. Thus the total mass of the particle would arise from
its coupling to the field.  Such a program, however, was not
feasible, since the self-energy diverged linearly, with no
realistic compensation possible.  In terms of renormalization
theory, this implied an infinite ``bare" mechanical mass.  Since
gravitational interaction energy is negative on the Newtonian
level, it may be expected to provide compensation. Indeed a simple
argument yields a limit on the self-energy due to just such a
compensation.  Consider a {\it bare} mass $m_0$ distributed in a
sphere of radius $\e$. In the Newtonian limit, the total energy
({\it i.e.}, the clothed mass) $m$ is given by $m = m_0 -
\frac{1}{2}\, \g m_0\,\!^2/\e$. For sufficiently small $\e ,\; m$
could become zero and then negative. In general relativity, the
principle of equivalence states that it is the total energy that
interacts gravitationally and not just the bare mass. Thus, as the
interaction energy grows more negative, were a point reached where
the total energy vanished, there could be no further interaction
energy. Consequently, there can be no negative total energy, in
contrast to the negative infinite self-energy of Newtonian theory.
General relativity effectively replaces  $m_0$  by $m$ in the
interaction term: $m = m_0 - \frac{1}{2}\, \g m^2/\e$. Solving for
$m$ yields $m = \g^{-1} [- \e + \e^2 + 2\g m_0\e)^{1/2}]$, which
shows that $m\rightarrow 0$ as $\e \rightarrow 0$.

More interesting is the fact that the gravitational interactions
produce a natural cutoff for the Coulomb self-energy of a point
charge.  Here the self-mass resides in the Coulomb field
$\frac{1}{2} \, \int (e/4\pi r^2)^2 \: d^3{\bf r}$. By the general
argument above on gravitational compensation, one expects that the
Coulomb energy near the origin (which is, in fact, ``denser" than
the neutral particle's $\d$-function distribution) will have a
very strong gravitational self-interaction, resulting in a
vanishing {\it total} contribution to the self-mass from the
region near the origin. Thus, the integral effectively extends
down only to some radius $a$, yielding $m_{{\rm EM}} = \frac{1}{2}
\: \int^\infty_a \: (e/4\pi r^2 )^2 d^3r = (e^2/4\pi )/2a$. We can
determine this effective flat-space cutoff $a$ by the same
equivalence principle argument. Without the gravitational
contribution, the mass is $m_0 + \frac{1}{2}(e^2/4\pi\e )$, so the
total clothed mass is determined by the equation $m = m_0 +
\frac{1}{2}\, e^2/4\pi\e - \frac{1}{2} \, \g m^2 /\e$. This yields
\be%7.1
m = \g^{-1} \{ - \e + [\e^2 + 2m_0\e\g + (e^2/4\pi )\g ]^{1/2}
\}\; .
 \ee
  We will see below that this formula is a rigorous
consequence of the field equations.  In the limit $\e \rightarrow
0$ , we have $m = (e^2/4\pi )^{1/2} \g ^{-1/2}$.  The bare
mechanical mass $m_0$ again does not contribute to the clothed
mass. Our result is then equivalent to a cutoff $a = \frac{1}{2}\,
(e^2/4\pi )^{1/2} \g ^{1/2}$  on the flat-space Coulomb integral.

\renewcommand{\theequation}{7.\arabic{equation}}
\setcounter{equation}{2}

{\bf 7.2. {\it Calculation of charged and neutral particle
self-energies from initial data.}}  We begin the self-energy
calculation, then, by obtaining a solution of the field equations
for a pure one-particle state, {\it i.e.}, a state containing no
independent excitations of the gravitational or electromagnetic
fields in the rest frame.  This requires that $g_{ij}\,\!^{TT} =
\pi^{ijTT} = A^{iT} = {\cal E}^{iT} = p_i = 0$ on the $t$ = const
surface where the energy is being computed. With the coordinate
conditions of (4.22), we are therefore dealing with the
time-symmetric situation $\pi^{ij} = 0$. According to (4.23),
then, the metric is isotropic; it is convenient to write it as
$g_{ij} = \chi^4 (r) \; \d_{ij}$.  From Section 7-5, the energy is
the coefficient of $1/(32\pi r)$ in the asymptotic form of $\chi
(r)$. The field equation determining $g^T$ and hence $\chi$ is
(6.6a). One has\footnote{Professor L.N. Cooper has pointed out to
us that (7.2a) with $e = 0 \; [-8\chi\nabla^2\chi = \rho_0 (r)$,
where $\rho_0$ is any bare mass distribution] can be obtained by
simple equivalence principle arguments starting from Newtonian
theory. The Poisson equation $\nabla^2\phi = 4\pi\g \rho_0 =
\frac{1}{4} \, \rho_0$ for the gravitational potential $\phi$,
must be corrected to include the particle's gravitational
self-energy $\frac{1}{2} \, \rho \phi$ as part of the source, {\it
i.e.}, $\nabla^2 \phi = \frac{1}{4} \rho \equiv  \frac{1}{4}
(\rho_0 + \frac{1}{2} \rho\phi )$. Eliminating $\rho$, one obtains
$\nabla^2\phi = \frac{1}{4} \, \rho_0 (1-\frac{1}{2}\, \phi
)^{-1}$. In terms of $\chi = 1 -  \frac{1}{2}\,\phi$, this is just
(7.2a) in the neutral case. It is interesting that this argument
yields the rigorous field equations in the frame (4.22) for this
situation. For the point electric charge, the same argument with
$\rho_0$ replaced by $\rho_0 + \frac{1}{2} \, \rho_e\phi_e$
($\phi_e$ being the electrostatic potential) leads to an equation
for $\phi$ which, while different from (7.2a), does yield the
correct total energy.}
$$%7.2a
\sqrt{g}\;\, ^3\!R = -8\chi\nabla^2\chi = m_0 \: \d^3({\bf r}) +
\textstyle{\frac{1}{2}} \:\chi^{-2}{\cal E}^{iL}{\cal E}^{iL}
\eqno{(7.2a)}
$$
and the electric field {\bf E}$^L$ by (6.3b) is
\be%7.3
{\cal E}^{iL} = (-e/4\pi r)_{,i} \; .
 \ee

\renewcommand{\theequation}{7.\arabic{equation}}
\setcounter{equation}{3}

A formal solution of (7.2a) may be found by setting $\chi^2 =
\psi^2 - \phi^2$  (Misner and Wheeler, 1957). One finds
$$%7.2b
- 8\chi \nabla^2\chi = 8 (\phi\nabla^2\phi - \psi\nabla^2\psi ) +
8 (\psi^2 - \phi^2 )^{-1} (\phi\nabla\psi - \psi\nabla\phi )^2
$$
$$
= m_0 \: \d^3({\bf r})+ \textstyle{\frac{1}{2}} \, ( \psi^2 -
\phi^2 ) ^{-1} ({\bf E}^L)^2 \; . ~~~~~~~~~ \eqno{(7.2b)}
$$
If one then makes the assumption
 \be%7.4
 {\bf E}^L = 4 (\phi\nabla\psi - \psi\nabla\phi )
 \ee
with $\phi = e/16\pi r$ and $(1 + m/32\pi r) = \psi$, (7.4)
correctly reproduces (7.3). Equation (7.2b) then determines the
total mass $m$ to be the $\e \rightarrow 0$ limit of (7.1):
 \be%7.5
 m = \lim_{\e\rightarrow 0} 16\pi \{ - \e + [\e^2 + (e/8\pi )^2 +
 m_0\e / 8\pi ]^{1/2}\} \; .
 \ee
In (7.5) the parameter $\e$ has been introduced by setting $\d^3
({\bf r})/r$ equal to $\d^3 ({\bf r})/\e$, and is thus essentially
the ``radius" of the $\d$-function. [This interpretation of the
$\d$-function corresponds to viewing $\d^3 ({\bf r})$  as the
limit of a shell distribution $\d (r - \e )/4\pi r^2$ of radius
$\e$. In V, it is shown that the results of this section are, in
fact, independent of the model chosen for $\d^3({\bf r})$ in the
point limit.] The result (7.5) is that $m = 2|e|$ and hence the
total mass is finite and independent of the bare mechanical mass.
The gravitationally renormalized electrostatic self-energy is now
finite.  The analysis also points out that mass only arises if a
particle has nongravitational interaction with a field of finite
range.\footnote{It is interesting to note that, even though $m =
0$ for the neutral particle, this does not imply that space is
everywhere flat.  The metric is indeed flat for $r > \e$, but
rises steeply in the interior.  For example, $\int^\infty_0 d^3r
\: \sqrt{^3g} \;^3\!R = m_0$, which shows that space is curved at
the origin.} For example, an electrically neutral particle coupled
to a Yukawa field would acquire a mass by virtue of {\it this}
coupling.

A solution may also be obtained for the case of two particles of
like charge. The results are consistent with those obtained above
for one particle (see V).  The energy is just $2(|e_1| + |e_2|)$,
that is, just the sum of the individual masses, independent of
both the mechanical masses and the interparticle separation
$r_{12}$. The absence of an interaction term can be traced to the
cancellation of the Newtonian attraction of the masses with the
Coulomb repulsion of the like charges.\footnote{This cancellation
implies that the solution should be static (since there is no
initial potential between the particles).  This has indeed been
shown to be the case for $m_{1,2} = 2|e_{1,2}|$ by Papapetrou
(1947).} This also checks with the result expected at large
separation for particles whose mass and charge are related by
$m_{1,2} = (e^2\,\!_{1,2}/4\pi\g )^{1/2}$. For opposite charges,
such a cancellation should not occur. However, we have not been
able to obtain a rigorous solution for a pure particle state of
two opposite charges.

All the usual Newtonian theory results may be obtained from (7.5)
or from the corresponding equation [(3.8) of V] for two bodies in
the appropriate limit. This ``dilute" limit consists in regarding
$m_0$ and $e$ small compared to $\e$ before passing to $\e = 0$,
thus preventing the strong (non-Newtonian) gravitational
interactions. Alternately, the ``dilute" limit represents a
perturbation expansion of the theory in powers of the coupling
constant $\g$. From (7.1) one easily sees that
 \be%7.6
 E \sim m_0 + (e^2/4\pi - \g m_0\,\!^2)/2\e + O(1/\e^2)\; .
 \ee
[For two bodies, the Newtonian and Coulomb interaction terms
$(e_1e_2/4\pi - \g (m_0)_1 (m_0)_2)/r_{12}$  also appear.] Since
$m_0$ and $e$ are unrelated, there is no cancellation in the
self-energy term of (7.6). In fact, the perturbation expansion
consists of an infinite series of more and more divergent terms.
The inapplicability of such an expansion is shown by the finite
rigorous answer $m = 2|e|$.  If one attempted to use standard
perturbation renormalization techniques, one would indeed find
that the theory was unrenormalizable (see V).

{\bf 7.3. {\it Determination of full metric due to a charge.
Stability of the particle. Vanishing of self-stress.}}  We have
been able to find the self energy purely from the initial value
(constraint) equations. The full solution of the problem consists
also in the specification of the $N,\; N_i(g_{0\m})$ for our frame
as well as the time development of the system (fuller details are
in VA). The equations determining $N$ and $N_i$ are obtained by
taking the time derivatives of the coordinate conditions
(4.22c,d). These are basically linear combinations of the
right-hand sides of (3.15a,b) [including, of course, the coupling
term from (6.7)]. Fortunately, the initial values and the
coordinate conditions for our problem reduce these equations to
quite simple form.  At the initial time, (3.15a) becomes
\be%7.7
\pa_0g_{ij} = N_{i|j} + N_{j|i}\; .
 \ee
Taking the time derivative of (4.22c), one obtains a homogeneous
equation for $N_i$ which has the solution
\be%7.8
N_i ({\bf r},0) = 0\; .
 \ee
The relevant part of (3.15b) is now just the $\pa_0\pi^{ii}$
equation. The coordinate condition (4.22d), the isotropic form of
$g_{ij}$, and the constraint equation (7.2) simplify this equation
to
\be%7.9
\pa_m (\chi^2 \: \pa_m N) = \textstyle{\frac{1}{4}} \, N \{
\chi^{-2}{\cal E}^{iL}{\cal E}^{iL} + m_0 \: \d^3({\bf r})\}\; .
 \ee
For the point particle, the solution for $N$ is just (see VA):
 \be%7.10
N = ( 1 + | e | /8\pi r)^{-1} = \chi^2 \; .
 \ee
Note that the total solution for $g_\mn$, is everywhere {\it
non-singular} (except, of course, at the particle).  In VA, it was
shown that this coordinate system led to a {\it completely}
non-singular metric  $g_\mn$ initially, for a spread-out
distribution of charge and mass with arbitrary $m_0$ and $\e$.
(This is in contrast to the standard Reissner-Nordstrom metric in
isotropic coordinates for which $N$ always has a singularity for
small enough $\e$.)

Further, the point charge is a stable object, since the equations
(7.8) and (7.10) for $g_\mn$ initially, lead to a static solution
$(\pa_tg_\mn = \pa_t\pi^{ij} = 0)$.  The cancellation of the
(repulsive) electrostatic self-forces by the gravitational ones is
made clear by examining the components of ${\fT}^{ij}$, the
total system's spatial stress density. The conservation
requirement on the total stress tensor ${\fT}^\mn$ implies that
${\fT}^{ij}\,\!_{,j} = - {\fT}^{0i}\,\!_{,0}=0$ for the
static case, where  ${\fT}^{ij}$ are the total system's spatial
stresses. Thus, in the notation of the orthogonal decomposition,
${\fT}^{ij}$  reduces to ${\fT}^{ij} = {\fT}^{ijTT} +
{\fT}^{ijT}$, since it is transverse. Spherical symmetry means
that  ${\fT}^{ijTT}$, vanishes, since no preferred transverse
direction can be distinguished. Hence, ${\fT}^{ij}$  has at
most one independent component which may be taken as  
${\fT}^{ii}$. For our static point solutions ${\fT}^{ii}$ (as
calculated from either the Landau-Lifshitz or Einstein
pseudotensors) vanishes everywhere for arbitrary $m_0$ and $e$. In
the rest frame, then, ${\fT}^\mn = \rho \: \d^\m_0 \: \d^\n_0$
where $\int \: d^3r \: \rho = m$, the clothed mass. In a moving
system, therefore, ${\fT}^\mn = \rho (dx^\m /d\t)(dx^\n/d\t )$
($\t$ is Lorentz proper time). The vanishing of ${\fT}^{ij}$ in
the rest frame, then, is necessary for the structure of the total
stress tensor to be that of a renormalized mass $m$. (This
requirement is stronger than the usual one that $P_\m = \int \:
d^3r {\fT}^0\,\!_\m$ transform like the energy-momentum of a
particle.\footnote{The standard static Reissner-Nordstrom solution
has ${\fT}^{ii} \neq 0$ but $\int \: d^3r \: {\fT}^{ii}=0$.
Thus it obeys the weaker condition on $P_\m$ but its total stress
tensor is not that of a particle. This is due to the presence of
the {\it phenomenological} pressure terms which were needed there
to stabilize the particle.}) Thus the point charge is a completely
stable object, without any {\it ad hoc} pressure (cohesive) terms
required, and its mass is completely determined by its field
interactions.

{\bf 7.4. {\it Comparison with standard Reissner-Nordstrom
solutions.}} It is interesting to compare our results with the
standard discussion of Schwarzschild (neutral) and
Riessner-Nordstrom (charged) solutions.  In these approaches, no
bare matter parameters were ever introduced, so that the
self-energy problem could not be formulated.  The source term used
in the standard treatment is the Maxwell stress tensor, together
with a matter source term of a perfect fluid with four-velocity
$U^\m$:
\be%7.11
T^\mn = (\rho_{00} + p_0) U^\m U^\n + p_0 g^\mn \; .
 \ee
The scalar ``proper rest-mass density" $\rho_{00}$ is, as can be seen
from this equation, the mass density in the locally inertial
rest-frame, while $p_0$ is the pressure in this frame (see, for
example, M\o ller,  1952).  In our treatment, on the other hand,
the matter was introduced dynamically rather than as an externally
prescribed source.  This was accomplished by including a
Lagrangian for the particle. Comparing (7.11) with the source
terms in our equations,\footnote{In VA, the comparison with (7.11)
was carried out in terms of a ``dust" model of the particle, that
is, with a Lagrangian describing the dynamics of a continuous
distribution of matter in gravitational interaction, and without
phenomenological pressure forces.} one finds that $\rho_{00} =
m_0\chi\,\!^{-6}\:\d^3 ({\bf r})$, which shows that $\rho_{00}$
includes ``clothing" effects through the $\chi^{-6}$-factor. The
pressure terms in the usual source tensor (7.11) are introduced
expressly in order to make the distribution stable.  Since a
pressure term summarizes the presence of other forces (not being
treated dynamically), these forces would contribute to the
clothing of the original mechanical mass and hence the parameter
$m_0$ would now represent this original mechanical mass {\it plus}
the clothing due to the forces giving rise to the pressure term.
In our analysis, where no phenomenological pressure terms are
invoked, one cannot at the outset ask that a solution be static.
In fact, extended initial distributions of matter and charge are
not, in general, stable, although the point limit does have this
property.

\noindent{\bf 8. Outlook}

\renewcommand{\theequation}{8.\arabic{equation}}
\setcounter{equation}{0}

{\bf 8.1. {\it Discussion of quantization.}} In this survey, we
have seen that general relativity may be viewed as an ordinary
classical field, once the meaning of the coordinate invariance has
been analyzed. It was then possible to treat the dynamics of the
gravitational field according to techniques common to classical
field theory. In this way, many physical properties could be
directly understood from their counterparts in, say,
electrodynamics.  For example, the canonical formalism provided a
unique definition of energy and gravitational radiation.  The
gravitational field, of course, also has aspects not found
elsewhere.  In particular, its sources are the total energies of
{\it all} other fields.  The attractive, static interaction, part
of the total energy provided the possibility of a compensating
effect on the flat-space self-energies of other fields.  Thus, a
stable classical point electron of finite mass exists when
gravitation is included.

A realistic elementary particle theory must, of course, be
formulated in quantum terms to determine whether the quantum
self-mass is also finite.  A full quantum treatment includes the
quantization of the dynamical modes of the gravitational field,
and not only of the effects arising due to a $q$-number source.
Since a complete set of Poisson bracket relations among the
classical canonical variables has been obtained, a correspondence
principle quantization may be performed immediately by
transcribing the P.B.'s into commutators.  Such a quantization may
indeed be valid when appropriate care of the ordering of the
canonical operators in the non-linear parts of the Hamiltonian is
taken.  However, certain consistency criteria proper to quantum
theory and extending beyond this possible ambiguity must be
examined.  The reduced system (involving only the canonical
formalism and its two degrees of freedom) does not represent the
full statement of general relativity: one also has equations to
determine $g_{0\m}$, for example, which are not part of this
canonical theory as such.  Furthermore, one encounters these
variables as soon as one makes a Lorentz transformation from the
initial frame,\footnote{In electrodynamics one similarly has that
the gauge functions ({\it i.e.}, the scalar potential and
longitudinal part of the vector potential) mix with the dynamical
variables ({\it i.e.}, the transverse parts of the vector
potential) when a Lorentz transformation is made.} an operation
that must be allowed for any sensible quantum theory. Since the
equations defining $g_{0\m}$ are now quantum ones, one must
establish the consistency of the ordering of the Hamiltonian and
the $g_{0\m}$   equations in one Lorentz frame with those in
another frame. Finally, we saw classically that there was an
infinite number of {\it a priori} equivalent sets of simple
canonical variables with conserved Hamiltonians, each of which
could, of course, be used as a basis for such a quantum scheme;
quantum mechanically, however, the relation between these sets of
variables need no longer be one of a unitary transformation due to
the operator character of the variables. Hence these are no longer
{\it a priori} equivalent starting points for quantization.  The
classical canonical transformations among these sets will cause
the coordinates associated with one set to be functions of both
the canonical variables and coordinates of the other.  This would
lead to the phenomenon, in the quantum theory, of the coordinates
of one set being $q$-numbers when expressed in terms of the
variables of the other set,\footnote{A similar situation could
arise in particle mechanics with a transformation $\bar{t} = t +
f(p,q)$. Here one would not so transform, since a preferred time
coordinate $t$ has been decided upon; in gravitation, however, no
such single preferred coordinate frame exists in the interior, at
least classically.  Whether the quantum theory may force a
preferred frame through the consistency requirements is not
known.} as was discussed in II and IVA.

In view of the many ambiguities which could arise in an attempt to
quantize consistently at this level, it would seem more fruitful
to return to the Lagrangian in four-dimensional form, {\it i.e.},
${\fg}^\mn \, R_\mn [\G^a\,\!_{\l\rho}]$ and try to repeat
our reduction to canonical form within the framework of quantum
theory.  There, one can use the manifest Lorentz covariance of the
original Lagrangian as an aid in proving the Lorentz covariance of
the canonical quantum form that should arise.  Further, the
ordering ambiguities are drastically reduced: since at most cubic
terms enter in this Lagrangian, one can show easily that there is
a simple three-parameter family of available Hermitian quantum
Lagrangians, all of which are generally covariant.  These
different orderings of the Lagrangian differ from each other only
by double commutators, {\it i.e.}, by effects of order $\hbar^2$.
The basic requirement of consistency between the Lagrange and
Heisenberg equations of motion should single out one of these
forms, since commutators of Bose fields do not contribute to
Lagrange equations but presumably affect the Heisenberg ones. In
an investigation based on the four-dimensional form of the quantum
Lagrangian, one ultimately expects to arrive at canonical forms
very similar to those obtained here. Thus the classical results
should represent an excellent guide in formulating the quantum
theory.\footnote{It has been shown in I that, if the theory can be
quantized at all, it obeys Bose statistics (as intuitively
expected).}

{\bf 8.2. {\it Speculations on quantum self-energy problem.}}
Leaving aside these technical questions on the rigor of
gravitational quantization, however, one may speculate as to the
quantum effects on the self-energy problem.  The numerical value
obtained classically for the mass $m$ of a point particle with
electronic charge, $m = e/(4\pi\g )^{1/2} \simeq 10^{18}m_e$, is
much too large. Of course, one would not expect classical theory
to give correct numerical values for masses.  Any realistic
discussion of self-masses must be based on quantum theory.
However, if the effective flat-space cutoff $(a \sim [ e/(4\pi
)^{1/2} ]\g^{1/2} \sim  10^{-34}$ cm)   obtained in this paper
were to hold also in quantum theory,\footnote{An effective quantum
gravitational cutoff might, on dimensional grounds, be $\sim (\g
\hbar c^{-3})^{1/2} \sim 10^{-33}$ cm. This differs from the
classical $a$ merely by the factor $a^{1/2}\equiv (e^2/4\pi\hbar
c)^{1/2}$ and would not affect the discussion in the text.} the
numerical values for the mass and charge would be quite different.
For example, using such a cutoff in Landau's estimate (Landau,
1956) for the renormalized charge, one finds $e_r\,\!^2/4\pi
\approx 10^{-2}$ (essentially independent of the bare charge) as
noticed by Landau. Thus, Landau obtains
\bea%8.1
e_r\,\!^2 & = & e^2 [1 + (2/3\pi )\n (e^2/4\pi ) \ln \, \{(\hbar
/mc)/a\}]^{-1}\nonumber \\
& \approx & [(\n /12\pi^2) \ln \, \{(\hbar /mc)/a\}]^{-1}
 \eea
where $\n \approx 10$ is essentially the number of charged fields.
Equation (8.1) was obtained by summing the dominant terms in each
self-energy diagram.  With an effective cutoff of physical origin,
the usual objections to making an estimate of what otherwise is a
divergent series need no longer hold.  The methods of Landau also
yield a formula for the renormalized mass in terms of the bare
mass and the charge.\footnote{This formula, $m = m_0
(e^2/e_r\,\!^2 )^{9/4\n}$, does not make clear what relation
between $m, \; m_0$, and $e_r$ might be expected, since the
estimate of (8.1) fails to determine $e$ with sufficient
accuracy.} In the classical theory, we saw that the bare mass did
not enter. To what extent this is maintained in the quantum theory
({\it e.g.}, to what extent the bare mass distribution remains
effectively a $\d$-function) is not clear. It should also be
emphasized that we have assumed in the above discussion the
simplest possibility, that the gravitational effects on the
quantum theory can be summarized in terms of a cutoff of the type
considered by Landau.

Finally, the treatment given here has not touched on several major
topics of interest in the classical theory.  For example, the
problem of motion (Einstein--Infeld--Hoffman analysis),has not
been examined from the present viewpoint.  Here, it is hoped that
the canonical methods will shed further light on the manner in
which the coupling of particles with the various components of the
metric determines their motion.  A particular issue where these
techniques might be relevant is the question of radiation by the
moving masses.  Some further topics involve spaces with non-flat
boundary conditions, which arise in cosmology, as well as spaces
where the topology is not Euclidian.  It remains to be seen
whether an approach of this type can be applied to such
situations.
\newpage

\begin{center}
 {\bf REFERENCES}
 \end{center}

\noindent Arnowitt, R., and S.\ Deser, 1959, {\it Phys.\ Rev.},
{\bf 113}, 745 (I).\\
Arnowitt, R., S.\ Deser, and C.W.\ Misner:\\
\hspace*{.3in} 1959, {\it Phys.\ Rev.}, {\bf 116}, 1322 (II).\\
\hspace*{.3in} 1960, {\it Phys.\ Rev.}, {\bf 117}, 1595 (III).\\
\hspace*{.3in} 1960, {\it Nuov.\ Cim.}, {\bf 15}, 487.\\
\hspace*{.3in} 1960, {\it Phys.\ Rev.\ Lett.}, {\bf 4}, 375.\\
\hspace*{.3in} 1960, {\it Phys.\ Rev.}, {\bf 118}, 1100 (IV).\\
\hspace*{.3in} 1960, {\it J.\ Math.\ Phys.}, {\bf 1}, 434 (IIIA).\\
\hspace*{.3in} 1960, {\it Phys.\ Rev.}, {\bf 120}, 313 (V).\\
\hspace*{.3in} 1960, {\it Phys.\ Rev.}, {\bf 120}, 321 (VA).\\
\hspace*{.3in} 1960, {\it Ann.\ Phys.}, {\bf 11}, 116 (VB).\\
\hspace*{.3in} 1961, {\it Nuov.\ Cim.}, {\bf 19}, 668 (IVA).\\
\hspace*{.3in} 1961, {\it Phys.\ Rev.}, {\bf 121}, 1556 (IVB).\\
\hspace*{.3in} 1961, {\it Phys.\ Rev.}, {\bf 122}, 997 (IVC).\\
Brill, D., 1959, {\it Ann.\ Phys.}, {\bf 7}, 466.\\
Eisenhart, L.P., 1949, {\it Riemannian Geometry}, Princeton
University Press,
Princeton.\\
Lanczos, C., 1949, {\it The Variational Principles of Mechanics},
Toronto University Press.\\
Landau, L.D., 1955, in {\it Niels Bohr and the Development of
Modern Physics}, \\
 \hspace*{.5in} Pergamon Press, London. \\
Misner, C.W., and J.A.\ Wheeler, 1957, {\it Ann.  Phys.}, {\bf 2}, 592.\\
M\o ller, C., 1952, {\it The Theory of Relativity}, Oxford
University Press.\\
Papapetrou, A., 1947, {\it Proc.\ Roy.\  Irish Acad.}, {\bf 51A},
191.\\
Petrov, A.Z., 1954, {\it Sci.\ Not.\ Kazan State Univ.}, {\bf
114},
55. \\
Pirani, F.A.E., 1957, {\it Phys.\ Rev.}, {\bf 105}, 1089.\\
Schr\"{o}dinger, E., 1950, {\it Space-Time Structure}, Cambridge
University Press. \\
Schwinger, J., 1951, {\it Phys.\ Rev.}, {\bf 82}, 914. \\
--------- 1953, {\it Phys.\ Rev.}, {\bf 91}, 713.
\end{document}